\documentclass[aip,graphicx]{revtex4-1}
\usepackage[cp1251]{inputenc}
\usepackage[T2A]{fontenc}
\usepackage[english]{babel}
\usepackage{amssymb,latexsym,amsmath,amscd}
\usepackage{graphicx,color,framed}
\usepackage{xcolor}
\usepackage{siunitx} 
\usepackage{empheq}
\usepackage{amsfonts}
\usepackage[normalem]{ulem}
\graphicspath{{figures/}}

\begin{document}
\title{Electrochemistry meets polymer physics: polymerized ionic liquids on an electrified electrode}
\author{\firstname{Yury A.} \surname{Budkov}}
\email[]{ybudkov@hse.ru}
\affiliation{School of Applied Mathematics, HSE University, Tallinskaya st. 34, 123458 Moscow, Russia}
\affiliation{G.A. Krestov Institute of Solution Chemistry of the Russian Academy of Sciences, 153045, Akademicheskaya st. 1, Ivanovo, Russia}
\author{\firstname{Nikolai N.} \surname{Kalikin}}
\affiliation{School of Applied Mathematics, HSE University, Tallinskaya st. 34, 123458 Moscow, Russia}
\affiliation{G.A. Krestov Institute of Solution Chemistry of the Russian Academy of Sciences, 153045, Akademicheskaya st. 1, Ivanovo, Russia}
\author{Andrei L. Kolesnikov}
\affiliation{Institut f\"ur Nichtklassische Chemie e.V., Permoserstr. 15, 04318 Leipzig, Germany}
\begin{abstract}
Polymeric ionic liquids are emerging polyelectrolyte materials for modern electrochemical applications. In this paper, we propose a self-consistent field theory of the polymeric ionic liquid on a charged conductive electrode. Taking into account the conformation entropy of rather long polymerized cations within the Lifshitz theory and electrostatic and excluded volume interactions of ionic species within the mean-field approximation, we obtain a system of self-consistent field equations for the local electrostatic potential and average concentrations of monomeric units and counterions. We solve these equations in the linear approximation for the cases of a point-like charge and a flat infinite uniformly charged electrode immersed in a polymeric ionic liquid and derive analytical expressions for local ionic concentrations and electrostatic potential, and derive an analytical expression for the linear differential capacitance of the electric double layer. We also find a numerical solution to the self-consistent field equations for two types of boundary conditions for the local polymer concentration on the electrode, corresponding to the cases of the specific adsorption absence (indifferent surface) and strong short-range repulsion of the monomeric units near the charged surface (hard wall case). For both cases, we investigate the behavior of differential capacitance as a function of applied voltage for a pure polymeric ionic liquid and a polymeric ionic liquid dissolved in a polar organic solvent. We observe that the differential capacitance profile shape is strongly sensitive to the adopted boundary condition for the local polymer concentration on the electrode.
\end{abstract}
\maketitle
\section{Introduction}
Ionic liquids (ILs) at charged electrodes have already been quite thoroughly studied experimentally and theoretically due to their wide range of applications as "designer solvents" and "filling" for supercapacitors \cite{fedorov2014ionic}. Their polymeric forms (PIL) are a relatively new-found \cite{ohno1998,ohno2004,ohno2007} class of materials with a shallow theoretical background, although demonstrating a wide spectrum of interesting properties of both ionic liquids and polymeric systems, which can lead to their potential application as ionic and solid-state electrolytes, fuel cell membranes, dispersants, ordered nano- and mesostructures, sorbents, carbon precursors, gating material in field-effect transistors, systems for water treatment, osmotic energy conversion, solar modulation, capillary electrophoresis, electrochromatography, among others \cite{Yuan2011, Mecerreyes2011, Eftekhari2017, fadeeva2021imidazolium,chen2018,peltekoff2019,gupta2019, YukiKohno2015,Tang2014,gao2021polarization}.

The theoretical investigations on the charged polymers near charged interfaces are practically limited to the cases of polyelectrolyte adsorption modeling (see reviews \cite{Dobrynin2005,netz2003neutral,andelman2000polyelectrolyte} and references therein). It is instructive to discuss the existing theoretical models of polyelectrolyte solutions on a charged surface. One way to theoretically consider such systems is multi-Stern layer theory \cite{vanderschee1984,Papenhuijzen1985,evers1986} - a lattice model extending classical works \cite{roe1974,scheutjens1979statistical,Scheutjens1980,borisov1994polyelectrolyte,friedsam2005adsorption} on uncharged polymer adsorption modeling by incorporating the electrostatic free energy in the partition function of the system or modeling of a single polyelectrolyte chain adsorption on a charged surface. Another approach treats the electrostatic potential and the concentration of the polyelectrolyte as continuous functions \cite{varoqui1991conformation,varoqui1993structure,borukhov1995,chatellier1996adsorption,borukhov1998scaling,joanny1999,podgornik1993stretching,brilliantov2016generation}. Taking into account the free energy contributions of the monomeric unit excluded volume via the second virial coefficient, the entropy of small ions and electrostatic interactions and minimizing it with respect to the concentrations and the potential, the authors obtained a system of two differential equations. The first of them is a generalized Poisson-Boltzmann equation and the second one -- a generalization of the self-consistent field equation of neutral polymers. In work \cite{borukhov1995} the authors theoretically investigated the behavior of a polyelectrolyte solution confined between two charged surfaces for several cases of monomeric charge distributions. The authors described the system within the mean-field approach, introducing a system of coupled differential equations, consisting of a modified Poisson-Boltzmann equation for the electrostatic potential and a self-consistent field equation for the polymer order parameter (square root of monomeric units concentration). Previously, similar results for the smeared case of the monomeric charge distribution without accounting for the monomeric unit excluded volume were obtained in \cite{varoqui1991conformation,varoqui1993structure,bohmer1990weak}. A numerical calculation of the obtained combined equations allowed the authors to compare the monomer concentration profiles with different types of charge distributions and also to conclude that the surface charge drastically influences the amount of polymer confined between the plates. In order to continue their theoretical investigation of the polyelectrolyte behavior near a single charged surface, based on the obtained numerical results, the authors \cite{borukhov1998scaling} also derived expressions for the amount of the adsorbed polymers for high- and low-salt regimes. In the high-salt case, the authors obtained two limiting regimes dependent on the value of the charged monomeric unit fraction, i.e. polyelectrolyte strength.

A study of polyelectrolyte solution adsorption at the ideal charged interface within the mean-field theory \cite{chatellier1996adsorption} in the linear approximation led to the analytical results for the polymer concentration profiles, which were shown to oscillate in the case of low salt concentrations. Further development of the self-consistent field theory approach \cite{joanny1999}, based on the ground state dominance approximation for the polymer order parameter of the Gaussian chains and consideration of the small surface charge, which justifies the linearization of the self-consistent field equation for the electrostatic potential, allowed the author to describe the polyelectrolyte adsorption in two limiting regimes of low and high ionic strength. The model successfully described the surface charge overcompensation and the following inversion of the surface charge due to the polyelectrolyte adsorption.

A nonlocal density functional theory of the polyelectrolyte solution near a charged planar surface was developed in \cite{Li2006,Li2006a}. The account of the hard-sphere contribution within the fundamental measure theory, the chain connectivity via the thermodynamic perturbation theory, and electrostatic correlations allowed the authors to reach good agreement with the results of simulations for ion and polyion segment density distribution as well as polyionic surface excess at varying properties. They also studied the changes in the ion valence, polyion chain length, and difference of the ion and polyion sizes on the microscopic structure and mean electrostatic potential, while accurately capturing the charge inversion and layering effects.

Two theoretical papers \cite{Kumar2017,Kumar2020} have been published, where the investigation of the electrode-PIL interface by broadband dielectric spectroscopy with theoretical modeling, based on the Rayleighian dissipation function formalism, allowed the authors to conclude that at zero potential there is a preabsorbed layer that influences the results of the impedance and capacitance measuring. Interestingly, the steady-state capacitance curves do not show pronounced asymmetry, and the overall shape is qualitatively the same as that of the corresponding IL.

To sum up the above, there are a decent number of studies devoted to theoretical modeling of charged polymer adsorption onto a charged surface. However, to the best of our knowledge, there are no thermodynamic models of EDL on a PIL-conductive electrode interface similar to those formulated for the conventional monomeric ILs \cite{fedorov2014ionic,kornyshev2007double,budkov2018theory,wu2011classical,jiang2011density,forsman2011classical,goodwin2017mean,Maggs2016,may2019differential,cruz2019effect,bazant2011double,avni2020charge,budkovJPCC2021}, describing differential capacitance as a function of applied voltage. Moreover, most of the models discussed above deal with dilute or semi-dilute polyelectrolyte solutions, accounting for the excluded volume interactions via the second virial term.

In this paper, we propose a self-consistent field theory of the PIL condensed phase on a charged conductive electrode. Taking into account the conformation entropy of rather long polymerized cations, electrostatic and excluded volume interactions of ionic species, we will derive a system of self-consistent field equations for the local electrostatic potential and local concentrations of the monomeric units and counterions. We will analyze the behavior of the EDL differential capacitance as a function of applied voltage and compare it with the differential capacitance behavior predicted in Kornyshev's analytical mean-field theory of low-molecular-weight ILs with the same physical parameters.

\section{Self-consistent field theory for polymeric ionic liquids}
In this section, we will formulate a self-consistent field (SCF) theory of PIL electrostatically disturbed by some external charges fixed in space with the density $\rho_{ext}(\bold{r})$. Let us consider a PIL consisting of polymerized cations, whose monomeric units carry a charge $q>0$ and low-molecular-weight anions with a charge $-q$. We assume that a polymerization degree of the cations is very large ($N\gg 1$) and we can neglect the translation entropy of the mass center of the polymer chains. Thus, the grand thermodynamic potential of the PIL is 
\begin{equation}
\Omega=F_{el}+\Omega_{ref},
\end{equation}
where
\begin{equation}
F_{el}=\frac{\varepsilon\varepsilon_0}{2}\int d\bold{r}\psi(\bold{r})\Delta\psi(\bold{r})+\int d\bold{r}\rho(\bold{r})\psi(\bold{r})
\end{equation}
is the electrostatic energy in the mean-field approximation with the local charge density 
\begin{equation}
\rho(\bold{r})=\rho_{ext}(\bold{r})+q\left(n_{p}(\bold{r})-n_{c}(\bold{r})\right),
\end{equation}
where the first term is the local charge density of the external charges and the second term is the local charge density of the monomer segments and counterions; $\varepsilon$ is the dielectric permittivity of polymeric ionic liquid taking into account polarization of the medium,  $\varepsilon_0$ is the empty space permittivity, $\Delta$ is the Laplace operator. The grand thermodynamic potential of the reference polymer system can be written as follows 
\begin{equation}
\Omega_{ref}=F_{conf}+F_{vol}-\int d\bold{r}\left(\mu_{p}n_{p}(\bold{r})+\mu_{c}n_{c}(\bold{r})\right),
\end{equation}
where
\begin{equation}
F_{conf}[n_{p}]=-\frac{k_{B}Tb^2}{6}\int d\bold{r}n_{p}^{1/2}(\bold{r})\Delta n_{p}^{1/2}(\bold{r})
\end{equation}
is the Lifshitz conformation free energy \cite{lifshitz1969some,khokhlov1994statistical} of the flexible polymer chains with a bond length $b$; $k_{B}$ is the Boltzmann constant, $T$ is the temperature. The contribution of the volume interactions of monomeric units and counterions is
\begin{equation}
F_{vol}[n_{p},n_{c}]=\int d\bold{r}f(T,n_{p}(\bold{r}),n_{c}(\bold{r}))
\end{equation}
where $f=f(T,n_{p}(\bold{r}),n_{c}(\bold{r}))$ is the free energy density as a function of local concentrations, $n_{p,c}(\bold{r})$, of monomeric units and counterions, which can be estimated within the asymmetric lattice gas model \cite{sanchez1978statistical,Maggs2016}
\begin{equation}
f=\frac{k_{B}T}{v}\left(\phi_c\ln\phi_{c}+\left(1-\phi_{c}-\phi_{p}\right)\ln\left(1-\phi_{c}-\phi_{p}\right)\right),
\end{equation}
where $\phi_{c}=n_{c}v$ and $\phi_{p}=n_{p}v$ are the local volume fractions of the counterions and monomeric units, respectively, $v$ is the elementary cell volume related to the bond length by the condition natural for a lattice model, $v=b^3$; $\mu_{p}$ and $\mu_c$ are the bulk chemical potentials of the monomeric units and counterions, respectively. The Euler-Lagrange equations, $\delta\Omega/\delta n_{p,c}(\bold{r})=0$ and $\delta\Omega/\delta \psi(\bold{r})=0$, have the following form
\begin{empheq}[left=\empheqlbrace]{align}\nonumber
\label{scf_eq}
&\bar{\mu}_{c}(\bold{r})-q\psi(\bold{r})=\mu_{c}\\
&\bar{\mu}_{p}(\bold{r})-\frac{k_{B}Tb^2}{6n_{p}^{1/2}(\bold{r})}\Delta n_{p}^{1/2}(\bold{r})+q\psi(\bold{r})=\mu_{p}\\\nonumber
&-\varepsilon\varepsilon_0\Delta\psi(\bold{r})=\rho_{ext}(\bold{r})+q\left(n_{p}(\bold{r})-n_{c}(\bold{r})\right),
\end{empheq}
where 
\begin{equation}
\bar{\mu}_{c}=\frac{\partial f}{\partial n_{c}}=k_{B}T\ln\left(\frac{\phi_c}{1-\phi_c-\phi_p}\right),
\end{equation}
\begin{equation}
\bar{\mu}_{p}=\frac{\partial f}{\partial n_{p}}=-k_{B}T\ln\left(1-\phi_{c}-\phi_{p}\right)
\end{equation}
are the intrinsic chemical potentials of the monomeric units and counterions, respectively. Taking into account that in the bulk solution, where $\psi=0$, the local electroneutrality condition, $n_{p}=n_c=n_0$, is fulfilled, we obtain the following expressions for the bulk chemical potentials of the species 
\begin{equation}
\mu_c=k_{B}T\ln\left(\frac{\phi_0}{1-2\phi_0}\right),
\end{equation}
\begin{equation}
\mu_p=-k_{B}T\ln\left(1-2\phi_0\right),
\end{equation}
where $\phi_0=n_0v$ is the bulk volume fraction of the monomeric units and counterions. 

\section{Linear theory}
Unfortunately, system (\ref{scf_eq}) of SCF equations obtained above for the local concentrations of monomeric units and counterions, $n_{p,c}(\bold{r})$, and electrostatic potential, $\psi(\bold{r})$, cannot be, in general, analytically solved. However, in a similar way as in the classical Debye-Hueckel theory of simple electrolyte solutions, we can analyze these equations in the limit of weak electrostatic interactions, where the local concentrations of monomeric units and counterions only slightly deviate from the bulk value, $n_0$, i.e.
\begin{equation}
n_{p}=n_{0}+\bar{n}_{p},~n_{c}=n_{0}+\bar{n}_{c}.
\end{equation}
where we have introduced small perturbations of concentrations ($|\bar{n}_{p,c}|\ll n_0$). Thus, introducing the small dimensionless variables $x_p={\bar n_p}/{n_0}$, $\bar \psi={q\psi}/{k_BT}$, $x_c={\bar{n}_c}/{n_0}$, we can linearize SCF equations (\ref{scf_eq}) and, after some algebra, obtain the following system of linear equations
\begin{empheq}[left=\empheqlbrace]{align}\nonumber
\label{lin_scf}   
    &\Delta^2x_p-\lambda^2\Delta x_p+\frac{\chi^4}{4}x_p=-\frac{12 q}{\varepsilon\varepsilon_0 k_BTb^2(1-\phi_0)}\rho_{ext}\\
    &\bar\psi=\frac{b^2}{12}(1-\phi_0)\Delta x_p-\phi_0x_p\\
    &x_c=\frac{(1-2\phi_0)\bar\psi-\phi_0 x_{p}}{1-\phi_0}\nonumber,
\end{empheq}
where we have also introduced the auxiliary parameters 
\begin{equation}
\lambda^2=\frac{12\phi_0}{b^2(1-\phi_0)}+\frac{1-2\phi_0}{r_{D}^2(1-\phi_0)},~\chi^4=\frac{48}{r_{D}^2b^2}\frac{1+2\phi_0}{1-\phi_0},
\end{equation}
with the Debye radius, $r_{D}=\left({q^2n_0}/{\varepsilon\varepsilon_0 k_BT}\right)^{-1/2}$, which comes from the presence of the mobile counterions. In what follows, we will analyze the solution of the system of linear equations (\ref{lin_scf}) for two physically relevant cases: (A) a point-like charge and (B) a flat charged electrode, immersed in a PIL.

\subsection{Case of point-like charge}
Here we would like to solve eqs. (\ref{lin_scf}) for the case, when the polymeric ionic liquid is disturbed by the fixed point-like test charge $q_0$ placed at the origin. Taking into account that the charge density of the test point-like charge is 
\begin{equation}
\rho_{ext}(\mathbf{r})=q_0\delta(\mathbf{r}),
\end{equation}
where $\delta(\bold{r})$ is the Dirac delta-function, we obtain
\begin{equation}
\Delta^2 x_p-\lambda^2\Delta x_p+\frac{\chi^4}{4}x_p=-\frac{12 qq_0}{\varepsilon\varepsilon_0 k_BTb^2(1-\phi_0)}\delta(\mathbf{r}).
\end{equation}
Further, using the integral Fourier transformation
\begin{equation}
x_p(\mathbf{r})=\int \frac{d\mathbf{k}}{(2\pi)^3}e^{i\mathbf{k}\mathbf{r}}\tilde x_p(\mathbf{k}),
\end{equation}
we arrive at the following expression for the Fourier-image 
\begin{equation}
\tilde x_p(\mathbf{k})=-\frac{12 qq_0}{\varepsilon\varepsilon_0 k_BTb^2(1-\phi_0)}\frac{1}{k^4+\lambda^2k^2+\frac{\chi^4}{4}},
\end{equation}
We only consider here the physically relevant case, $\lambda < \chi$, which can be reformulated as the following condition for the Debye radius and bond length of the chain:
\begin{equation}
\label{cond_osc}
\frac{\sqrt{(1+2\phi_0)(1-\phi_0)}-1}{2\sqrt{3}\phi_0}<\frac{r_D}{b}<\frac{\sqrt{(1+2\phi_0)(1-\phi_0)}+1}{2\sqrt{3}\phi_0}.
\end{equation}
Condition (\ref{cond_osc}) covers the cases of pure PILs and PILs dissolved in a polar organic solvent like acetonitrile (see the physical parameters below). In this case, the inverse Fourier-transform yields
\begin{equation}
\label{xp}
x_p=a\frac{\exp\left[-\varkappa r\right]}{r}\sin{kr},
\end{equation}
where we have introduced the constants $\varkappa={\sqrt{\chi^2+\lambda^2}}/{2}$, $k={\sqrt{\chi^2-\lambda^2}}/{2}$ (see the discussion of their physical sense below) and coefficient
\begin{equation}
a=-\frac{3qq_0}{2\pi\varepsilon \varepsilon_0 k_BTb^2\varkappa k(1-\phi_0)}.
\end{equation}
The dimensionless electrostatic potential has the following form
\begin{equation}
\label{psi}
\bar\psi=\frac{\exp\left[-\varkappa r\right]}{r}(b_1\sin{kr}+b_2\cos{kr}),
\end{equation}
where
\begin{equation}
b_1=a\left(\theta-\phi_0\right),~b_2=-a\eta,
\end{equation}
where $\theta=\chi^2b^2(1-\phi_0)/24$ and $\eta=\varkappa k b^2 (1-\phi_0)/6$.

The perturbation of the cation concentration is determined by 
\begin{equation}
\label{xc}
x_c=\frac{\exp\left[-\varkappa r\right]}{r}(c_1\sin{kr}+c_2\cos{kr})
\end{equation}
with the coefficients
\begin{equation}
c_1=\frac{(1-2\phi_0)b_1-\phi_0 a}{1-\phi_0},
~c_2=\frac{1-2\phi_0}{1-\phi_0} b_2.
\end{equation}
It is interesting to note that for $\phi_0=0.5$ (there are no voids in the bulk liquid structure) we get $x_{p}=-x_c$ or $n_{p}+n_{c}=2n_0$. To the best of our knowledge, eqs. (\ref{xc}) and (\ref{xp}) are new analytical results. We would like to note that expression, similar to eq. (\ref{psi}), was obtained within the field theories of polyelectrolyte solutions in papers \cite{Borue1988,muthukumar1996double}.

We would like to note that the length $\varkappa^{-1}$ characterizes the spatial decay of the perturbations of the concentrations of monomeric units and counterions, and electrostatic potential relative to the bulk values, and is analogous to the Debye radius in simple electrolyte solutions. The length $k^{-1}$ can be considered as a measure of the $"$periodicity$"$ of positive and negative charge-rich domains, which is the result of the fine interplay between the conformation entropy of the polymer chains,  electrostatic interactions, and excluded volume interactions of monomeric units. We would like to mention that similar behavior of the electrostatic potential of the point-like charge has been recently predicted in salt solutions of quadrupolar molecules \cite{budkov2020statistical,budkov2019statistical,slavchov2014quadrupole} and ionic liquids with an account of the quadrupolar ionic clusters \cite{avni2020charge}.

\subsection{Case of a flat charged electrode}
Now, we would like to consider the above linear response theory for the case of a flat infinite uniformly charged electrode with the surface charge density $\sigma$. The electrode charge density is
\begin{equation}
\rho_{ext}(z)=\sigma\delta(z),
\end{equation}
so that the equation for the perturbation of monomeric unit concentration has the form
\begin{equation}
\label{flat_el_x_p}  
   x_p^{(IV)}(z)-\lambda^2x_p''(z)+\frac{\chi^4}{4}x_p(z)=-\frac{12 q\sigma}{\varepsilon\varepsilon_0 k_BTb^2(1-\phi_0)}\delta(z),
\end{equation}
where the superscript ${(IV)}$ denotes the fourth derivative. The boundary conditions can be written as follows
$x_p'(0)=0$, $x_p'''(0)=-{12 q\sigma}/{(\varepsilon\varepsilon_0 k_BTb^2(1-\phi_0))}$, $x_p(\infty)=x_p'(\infty)=0$,
where we adopt that there is no specific adsorption of polymer chains on the electrode surface, so that the boundary condition $n_{p}^{\prime}(0)=0$ is fulfilled \cite{netz2003neutral,andelman2000polyelectrolyte} and take into account the delta-function term in the right-hand side of eq. (\ref{flat_el_x_p}). The generalization of the theory for the case of specific adsorption of monomeric units on the electrode is straightforward ({see discussion below}). The solution of eq. (\ref{flat_el_x_p}) with the help of the above boundary condition results in  
\begin{equation}  
\label{xp2}
x_p(z)=e^{-\varkappa z}\left(f_1\cos(kz)+f_2\sin(kz)\right),
\end{equation}
with the following coefficients
\begin{equation}
f_1=-\frac{6 q\sigma}{\varkappa(\varkappa^2+k^2)\varepsilon\varepsilon_0 k_{B}Tb^2(1-\phi_0)},~f_2=-\frac{6 q\sigma}{k(\varkappa^2+k^2)\varepsilon\varepsilon_0 k_{B}Tb^2(1-\phi_0)}.
\end{equation}
The dimensionless electrostatic potential takes the following form 
\begin{equation}
\label{psi2}
\bar{\psi}=e^{-\varkappa z}\left(g_1\cos(kz)+g_2\sin(kz)\right),
\end{equation}
where 
\begin{equation}
g_1=\left(\theta-\phi_0\right)f_1-\eta f_2,
\end{equation}
\begin{equation}
g_2=\left(\theta-\phi_0\right)f_2+\eta f_1.
\end{equation}
The perturbation of the counterion concentration has the form
\begin{equation}
\label{xc2}
x_c=e^{-\varkappa z}\left(h_1\cos(kz)+h_2\sin(kz)\right),
\end{equation}
with the following coefficients linearly expressed via the ones written above
\begin{equation}
h_1=\frac{(1-2\phi_0)g_1-\phi_0f_1}{1-\phi_0},~h_2=\frac{(1-2\phi_0)g_2-\phi_0f_2}{1-\phi_0}.
\end{equation}
As in the above case of the point-like test charge, it is interesting to note that for $\phi_0=0.5$ we obtain $x_{p}=-x_{c}$ or $n_{c}(z)+n_{p}(z)=2n_{0}$.

Introducing the potential drop on the electrode $\psi_{0}=\psi(0)$, using eq. (\ref{psi}), after some algebraic calculations, we obtain the following expression for the surface charge density
\begin{equation}
\sigma=C_0\psi_0,
\end{equation}
with the linear differential capacitance of the EDL
\begin{equation}
\label{cap}
C_{0}=\frac{\varepsilon\varepsilon_0}{d},
\end{equation}
where 
\begin{equation}
\label{thikness}
d=\frac{1+\frac{\phi_0}{\theta}}{\sqrt{\lambda^2+\chi^2}}
\end{equation}
can be interpreted as effective thickness of the diffuse EDL. We would like to note that the thickness is proportional to determined above the effective screening length, $\varkappa^{-1}$. Eqs. (\ref{cap}) and (\ref{thikness}) are the main analytical result of this paper. It is interesting to note that in contrast to the case of low-molecular-weight ionic liquids described within the symmetric lattice gas model \cite{kornyshev2007double}, the differential capacitance, $C_0$, of the PILs in the linear theory depends on the molecular parameters of the ionic species, namely, the bond length, $b$, and the effective excluded volume of particles, $v$. In the limit of negligible excluded volume interactions ($v\to 0$) at constant bond length, $b$, we arrive at the expression
\begin{equation}
\label{GC_gen}
C_{0}=C_{GC}\sqrt{1+\frac{4\sqrt{3}r_D}{b}},
\end{equation}
which is an analog of the well-known linear Gouy-Chapman or Debye linear capacitance \cite{kornyshev2007double}, $C_{GC}=\varepsilon\varepsilon_0/r_{D}$, for low-molecular-weight electrolyte solutions and monomeric ionic liquids. Note that eq. (\ref{GC_gen}) should be valid for rather dilute polyelectrolyte solutions on a charged electrode in the theta-solvent regime, when the polymer chains behave as the ideal Gaussian coils \cite{khokhlov1994statistical}. In the region $r_{D}\ll b$, eq. (\ref{GC_gen}) transforms into the standard Gouy-Chapman expression, i.e. $d\simeq r_{D}$. However, when $r_{D}\gg b$, the electric double layer thickness is determined by $d\simeq 48^{-1/4}(r_{D}b)^{1/2}$. Note that a decrease in the bond length, $b$, as is seen from eq. (\ref{GC_gen}), increases the linear differential capacitance. Such behavior can be explained by the fact that the decrease in the bond length leads to higher linear charge density of the polymer chains and, thus, higher accumulated charge in the EDL.  

\section{Polymeric ionic liquid on a charged electrode: nonlinear regime}
\subsection{Statement of problem}
Now we would like to consider the case of a PIL on a flat charged electrode with a constant surface potential, $\psi_0$. The self-consistent field equations take the following simplified form 
\begin{empheq}[left=\empheqlbrace]{align}\nonumber
\label{scf_eq_2}
&\bar{\mu}_{p}-\frac{k_{B}Tb^2}{6\phi_{p}^{1/2}}(\phi_{p}^{1/2})^{\prime\prime}+q\psi=\mu_{p}\\
&\psi^{\prime\prime}=-\frac{q}{\varepsilon\varepsilon_0 v}\left(\phi_{p}-\phi_{c}\right)\\\nonumber
&\phi_c=(1-\phi_{p})\frac{e^{\frac{\mu_c+q\psi}{k_{B}T}}}{1+e^{\frac{\mu_c+q\psi}{k_{B}T}}}.
\end{empheq}
{In order to solve eqs. (\ref{scf_eq_2}), it is necessary to formulate the boundary conditions. Standard boundary conditions are applied for the electrostatic potential: $\psi(0)=\psi_{0},~\psi^{\prime}(\infty)=0$. The general boundary conditions for the polymer local volume fraction are $\phi_{p}^{\prime}(0)/\phi_{p}(0)=-2/\xi$, $\phi_{p}(\infty)=\phi_0$, where the so-called extrapolation length, $\xi$, is inversely proportional to the energy of short-range specific interaction between the monomer unit and the surface~\cite{andelman2000polyelectrolyte}. It is positive and of the order of a molecular size in the case of attractive polymer-surface interactions. The length $\xi$ is negative if the chains are repelled by the surface and it vanishes for an impenetrable surface (a hard wall), where $\phi_{p}(0)$=0. Below we consider the case of indifferent surface for which $1/\xi=0$ and, therefore, $\phi_{p}^{\prime}(0)=0$, i.e. when there is no specific adsorption of monomeric units onto the electrode and the case of a hard wall, when $\phi_{p}(0)=0$. In our opinion, the indifferent electrode regime corresponds most closely to the situation considered within the framework of Kornyshev's theory for monomeric ionic liquids. The hard wall case corresponds to that of strong (infinite) specific polymer-electrode repulsion at small distances.}

Solving eqs. (\ref{scf_eq_2}) at different potential drops, $\psi_0$, in two mentioned limiting regimes, we can calculate the surface charge density as $\sigma=-\varepsilon\varepsilon_0\psi^{\prime}(0)$ and then the differential capacitance, $C=d\sigma/d\psi_0$.

\subsection{Numerical results and discussions}

\begin{figure}[h]
\centering
\includegraphics[width=0.75\linewidth]{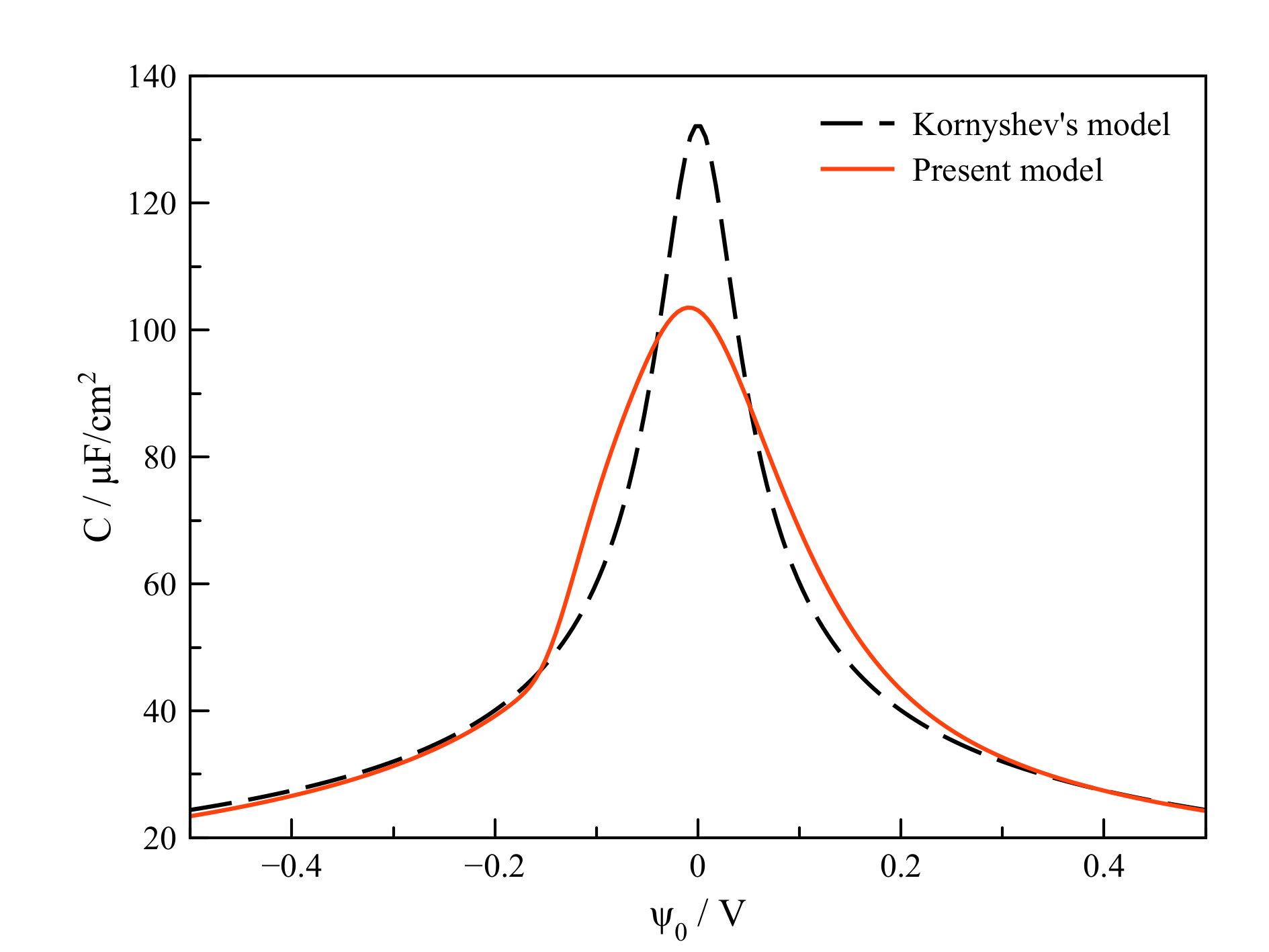}
\caption{Differential capacitance profile obtained numerically by present theory for pure PIL and differential capacitance predicted by the analytical mean-field theory for a monomeric IL (see eq. (\ref{Korn_cap})). The data are shown for $\phi_0=0.4$, $b=v^{1/3}=0.5~nm$, $T=298~K$, $\varepsilon=5$, $q=1.6\times 10^{-19}~C$.}
\label{fig1} 
\end{figure}

\textbf{Indifferent electrode case.} Solving numerically the self-consistent field equations (\ref{scf_eq_2}) {with the boundary condition $\phi_{p}^{\prime}(0)=0$}, we can calculate the differential capacitance as a function of applied voltage. Fig. \ref{fig1} shows a comparison of the differential capacitance profile of a PIL numerically calculated within the present theory and differential capacitance for monomeric IL determined by the following analytical expression derived by Kornyshev in ref. \cite{kornyshev2007double}
\begin{equation}
\label{Korn_cap}
C=\frac{\left({2\beta\varepsilon\varepsilon_0n_0^2vq^2}\right)^{1/2}|\sinh{\beta q\psi_0}|}{\left(1+4\phi_0\sinh^2\left(\frac{\beta q\psi_0}{2}\right)\right)\sqrt{\ln\left(1+4\phi_0\sinh^2\left(\frac{\beta q\psi_0}{2}\right)\right)}}.
\end{equation}

\begin{figure}[h]
\centering
\includegraphics[width=0.75\linewidth]{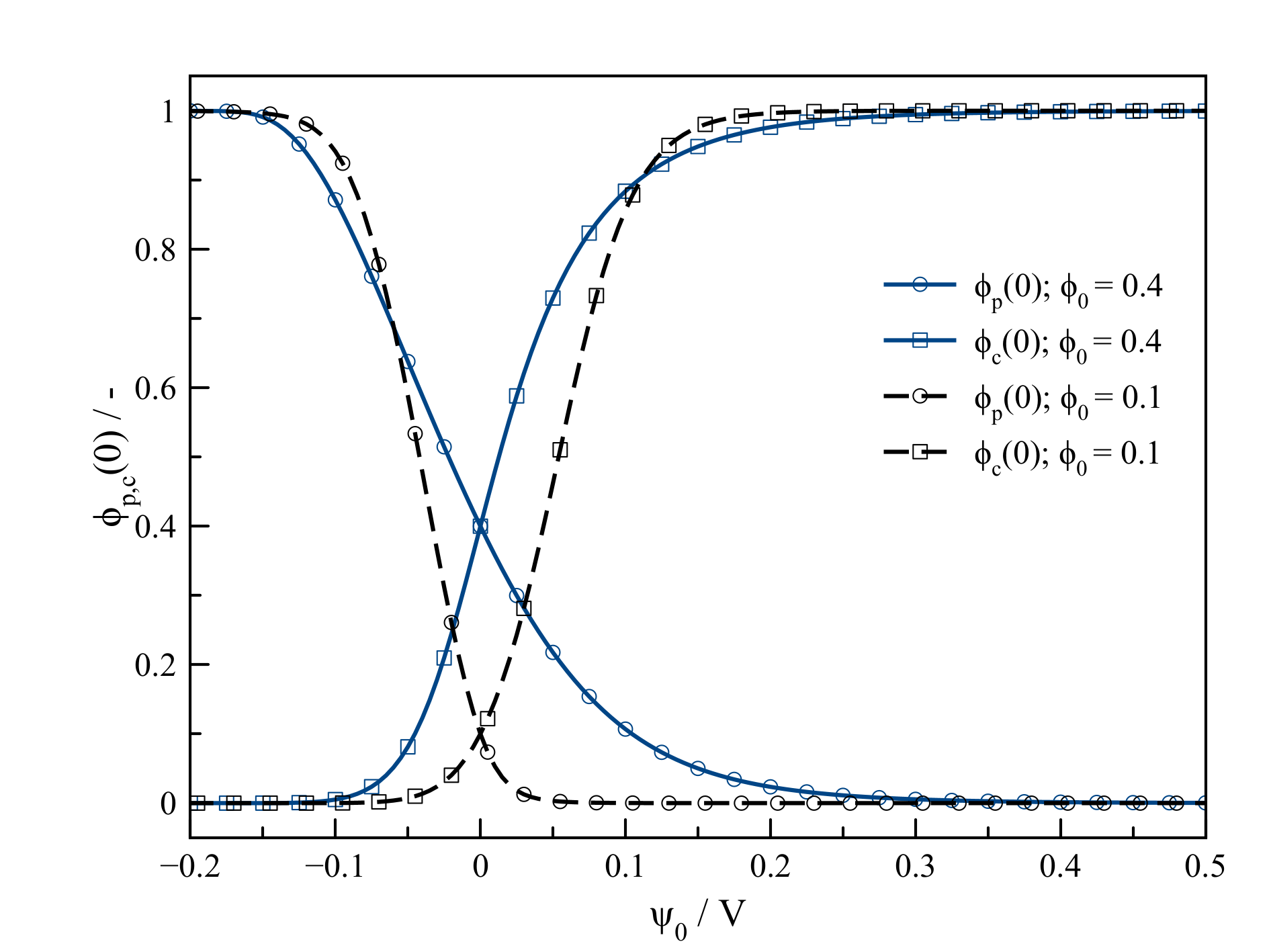}
\caption{Local volume fractions of monomeric units and counterions on the electrode as functions of the potential drop for the cases of a pure polymeric ionic liquid and and a polymeric ionic liquid solution. The parameters for the pure polymeric ionic liquid (blue lines) are $\phi_0=0.4$, $b=v^{1/3}=0.5~nm$, $T=298~K$, $\varepsilon=5$. The parameters for the polymeric ionic liquid solution (black lines) are $\phi_0=0.1$, $b=v^{1/3}=0.5~nm$, $T=298~K$, $\varepsilon=40$, $q=1.6\times 10^{-19}~C$.}
\label{fig2} 
\end{figure}

As is seen, as in case of a monomeric IL, the PIL differential capacitance profile has a quite symmetric bell shape (see also the discussion below). It is interesting to note that at sufficiently large potential drops, as in case of a monomeric IL, the differential capacitance monotonically decreases as $C\sim |\psi_0|^{-0.5}$. Moreover, at rather large positive potential drops, the differential capacitance profile coincides with the one predicted by Kornyshev's theory. At sufficiently large positive surface potentials, the monomeric units are fully expelled from the electrode surface by the counterions (see Fig. \ref{fig2}). Thus, the EDL has the same local structure as in the case of monomeric ILs, which explains the correspondence to Kornyshev's theory. At rather large negative surface potentials, the differential capacitance values are also close to those predicted by the Kornyshev mean-field theory. This can be explained by the fact that at rather large local volume fractions of the monomeric units near the electrode ($\phi_{p}\sim 1$), occurring at sufficiently large potential drops, there is no difference between the tied charged monomeric units and freely moving ions from the entropic standpoint -- the translational entropy in both cases is negligible.

\begin{figure}[h]
\centering
\includegraphics[width=0.75\linewidth]{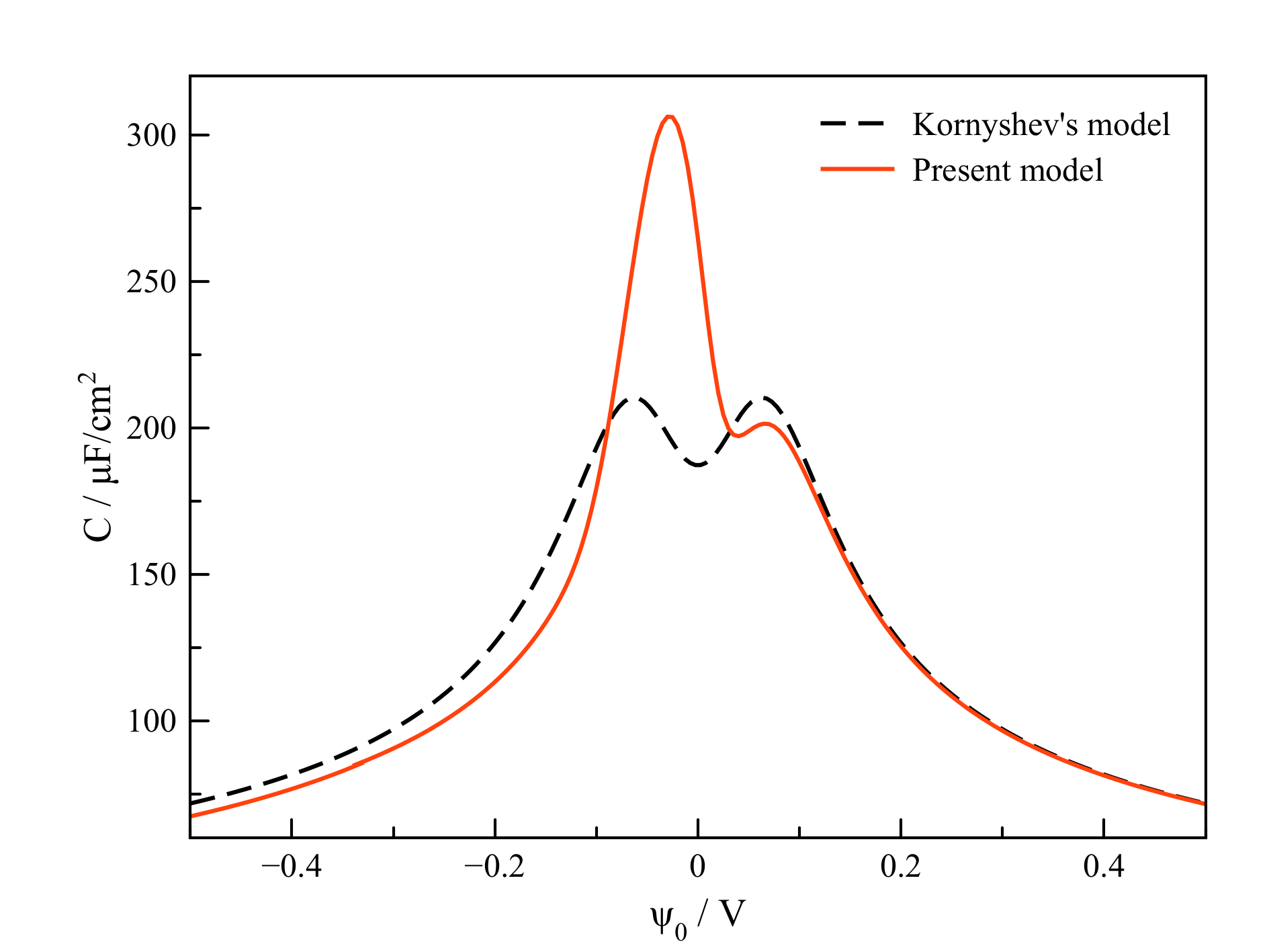}
\caption{Differential capacitance profiles predicted by the present theory for a PIL solution and predicted by Kornyshev's mean-field theory for a simple electrolyte solution (eq. (\ref{Korn_cap})). The data are shown for $\phi_0=0.1$, $b=v^{1/3}=0.5~nm$, $T=298~K$, $\varepsilon=40$, $q=1.6\times 10^{-19}~C$.}
\label{fig3} 
\end{figure}

Now it is instructive to discuss the differential capacitance profile realized for PILs dissolved in a highly polar organic solvent with the dielectric permittivity $\varepsilon\approx 40$. We assume the bulk volume fraction of the monomeric units to be $\phi_0=0.1$. Fig. \ref{fig3} demonstrates a comparison between the differential capacitance profile predicted by the present theory and Kornyshev's mean-field theory for the cases of a PIL solution and a simple electrolyte solution, respectively. Unlike the case of a pure PIL discussed above, where we obtained quite a symmetric bell-shaped capacitance profile, the differential capacitance profile for the PIL solution case has two strongly asymmetric peaks. The higher one is located in the region of negative voltages, whereas the less pronounced peak is in the region of positive voltages. As is seen, such capacitance behavior drastically differs from the behavior observed in simple electrolyte solutions, predicted by Kornyshev's theory for the same physical parameters. In particular, at negative voltages, when there are more positively charged monomeric units than counterions in the vicinity of the electrode surface, we obtain much higher differential capacitance values than those realized for the low-molecular-weight electrolyte solution. Such abnormal differential capacitance behavior can be explained by the fact that an increase in the negative voltage value leads to a rather dramatic increase in the local volume fraction of monomeric units on the electrode, $\phi_{p}(0)$, and the same decrease in the local volume fraction of the counterions, $\phi_{c}(0)$ (see Fig. \ref{fig2}). However, an increase in the positive voltage leads to a slower growth in the local volume fraction of the counterions and decrease in the local volume fraction of monomeric units. Such difference is associated with the fact that in the case of very large polymerization degree the long polymer chains have zero translation entropy and, therefore, can be more easily attracted to the negatively charged electrode than the mobile counterions to the positively charged electrode. We would like to note that in case of a PIL solution, as in case of a pure PIL, the differential capacitance at sufficiently high potential drop values is described by the same power law, $C\sim |\psi_0|^{-0.5}$.

\begin{figure}[h]
\centering
\includegraphics[width=0.75\linewidth]{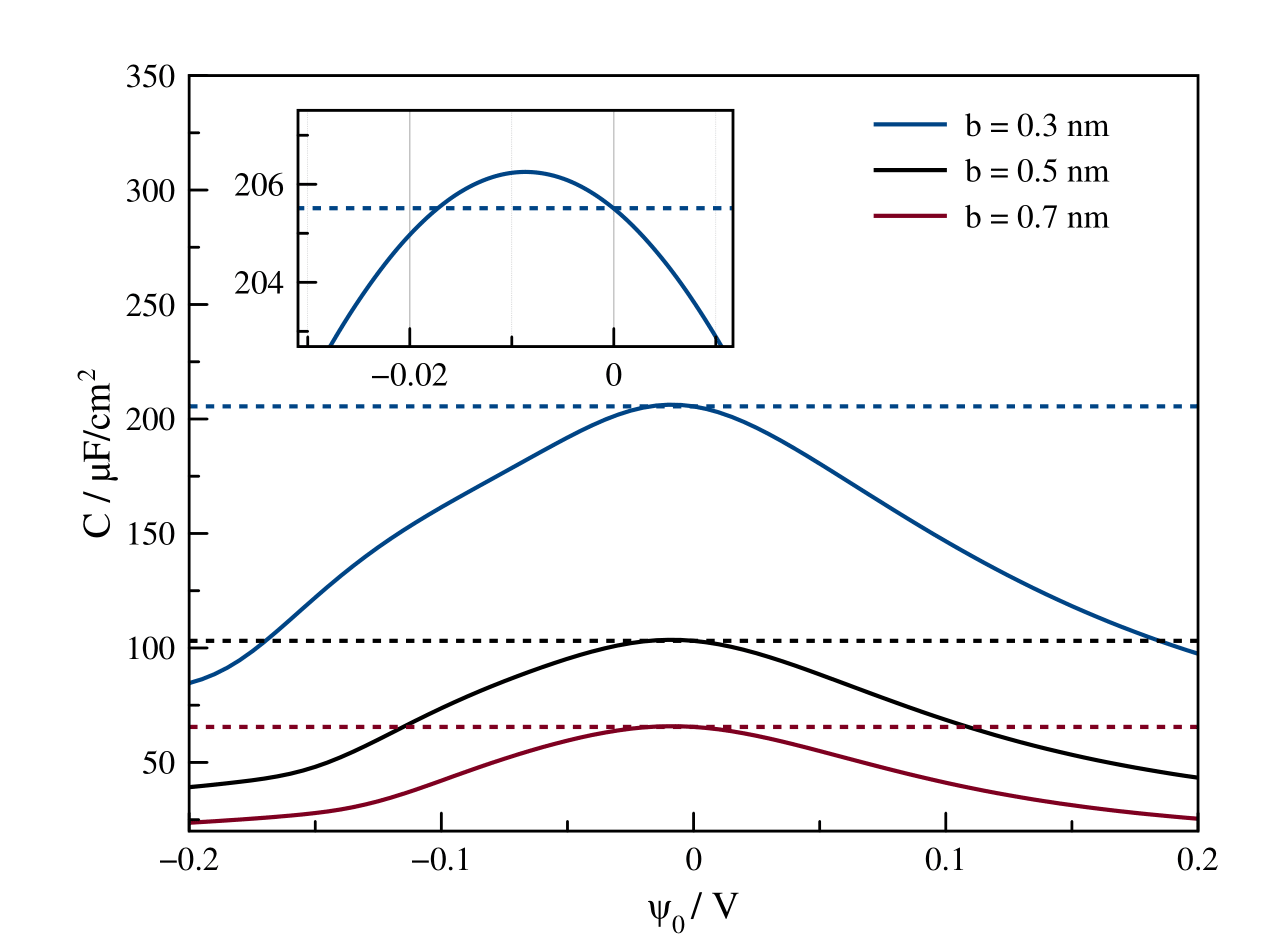}
\caption{Differential capacitance profiles of a pure PIL on a charged electrode plotted for different bond lengths, $b=v^{1/3}$. The horizontal dashed lines determine the linear capacitance (see eq. (\ref{cap})). The data are shown for $\phi_0=0.4$, $T=298~K$, $\varepsilon=5$, $q=1.6\times 10^{-19}~C$.}
\label{fig4} 
\end{figure}

Now we turn to the discussion of how a variation in the bond length, $b=v^{1/3}$, influences the differential capacitance. As is seen in Fig. \ref{fig4}, an increase in the bond length results in a monotonic decrease in the differential capacitance at all potential drops. Such differential capacitance behavior can be explained by the fact that the EDL accumulates a smaller charge with the increase in the bond length and excluded volume of particles due to stronger steric interactions of the ionic species in the EDL. It is also interesting to note that the differential capacitance maximum (see also Fig. \ref{fig1}) is slightly shifted to the negative potential drop values. Thus the predicted bell-shaped capacitance profile can be considered a symmetric one with good accuracy. This is in agreement with the conclusion of paper \cite{Kumar2020} discussed in the Introduction. Note that the same behavior with an increase in the bond length is observed in the PIL solution discussed above -- differential capacitance increases with the bond length. 

\begin{figure}[h]
\centering
\includegraphics[width=0.6\linewidth]{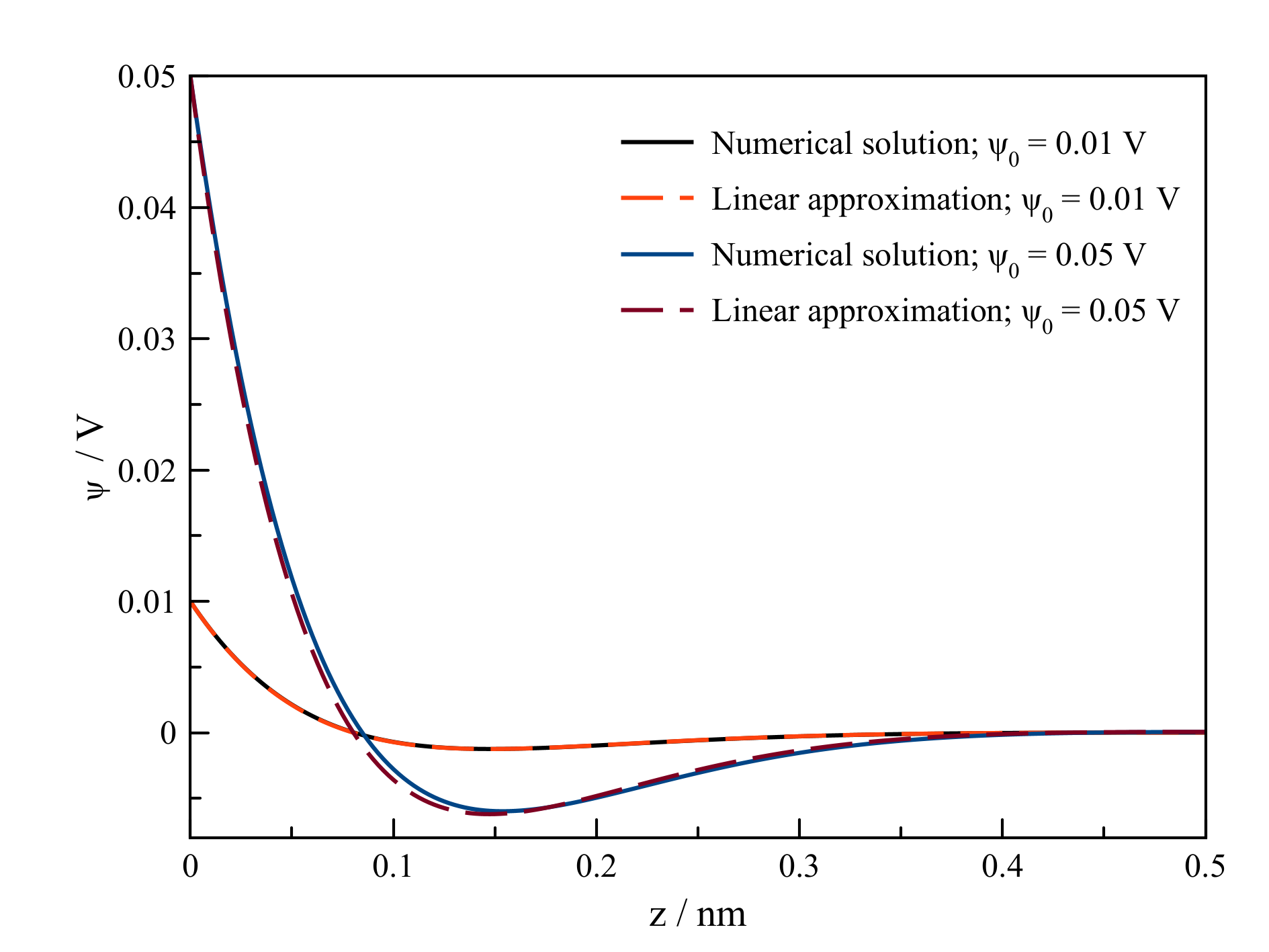}
\includegraphics[width=0.6\linewidth]{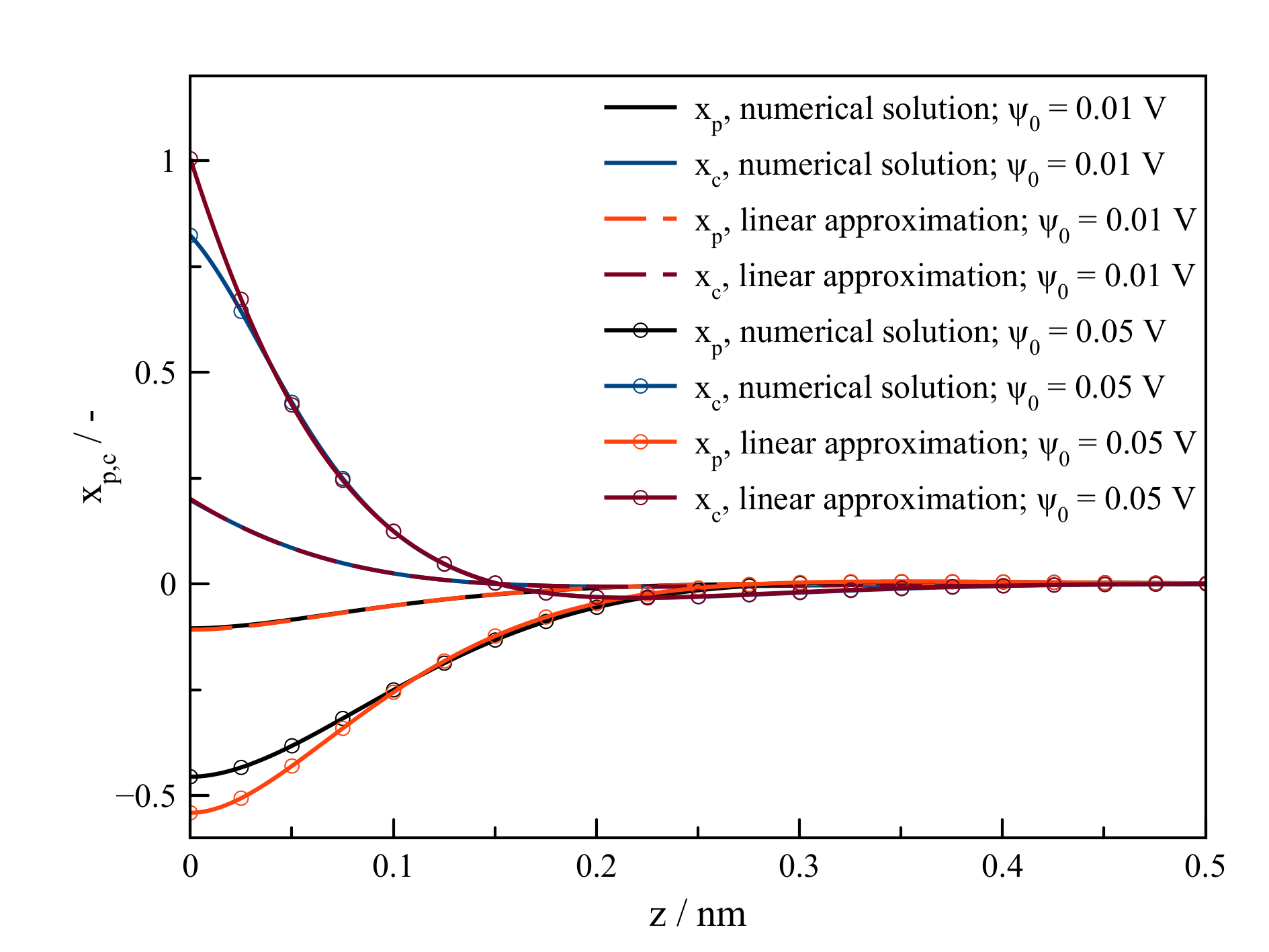}
\caption{Electrostatic potential (top) and concentration (bottom) profiles calculated within the linear analytical theory and nonlinear theory for positive potential drops. The data are shown for $\phi_0=0.4$, $b=v^{1/3}=0.5~nm$, $T=298~K$, $\varepsilon=5$, $q=1.6\times 10^{-19}~C$. The local concentrations are determined via $x_{p,c}(z)=(n_{p,c}(z)-n_0)/n_0$.}
\label{fig5} 
\end{figure}

\begin{figure}[h]
\centering
\includegraphics[width=0.6\linewidth]{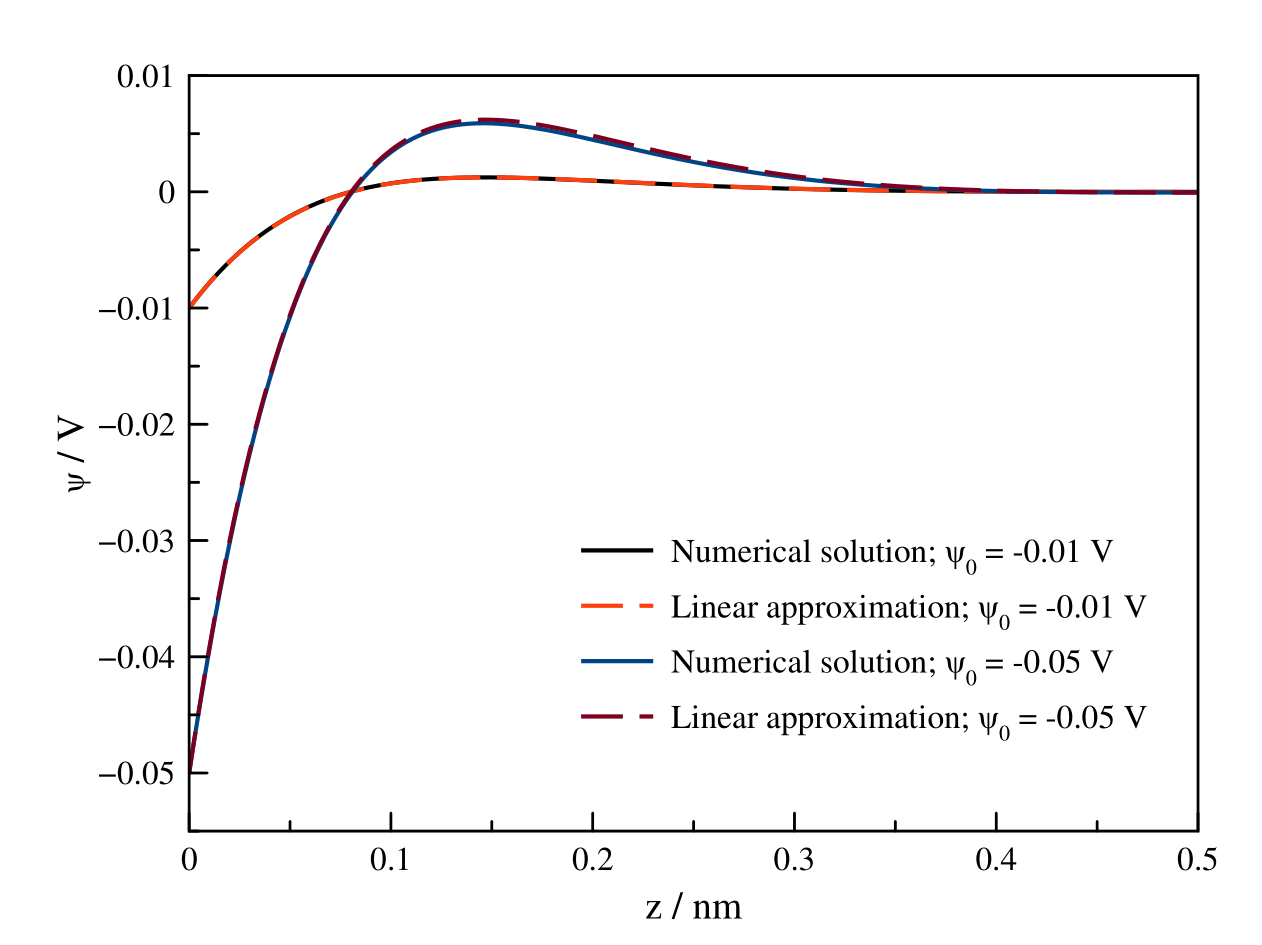}
\includegraphics[width=0.6\linewidth]{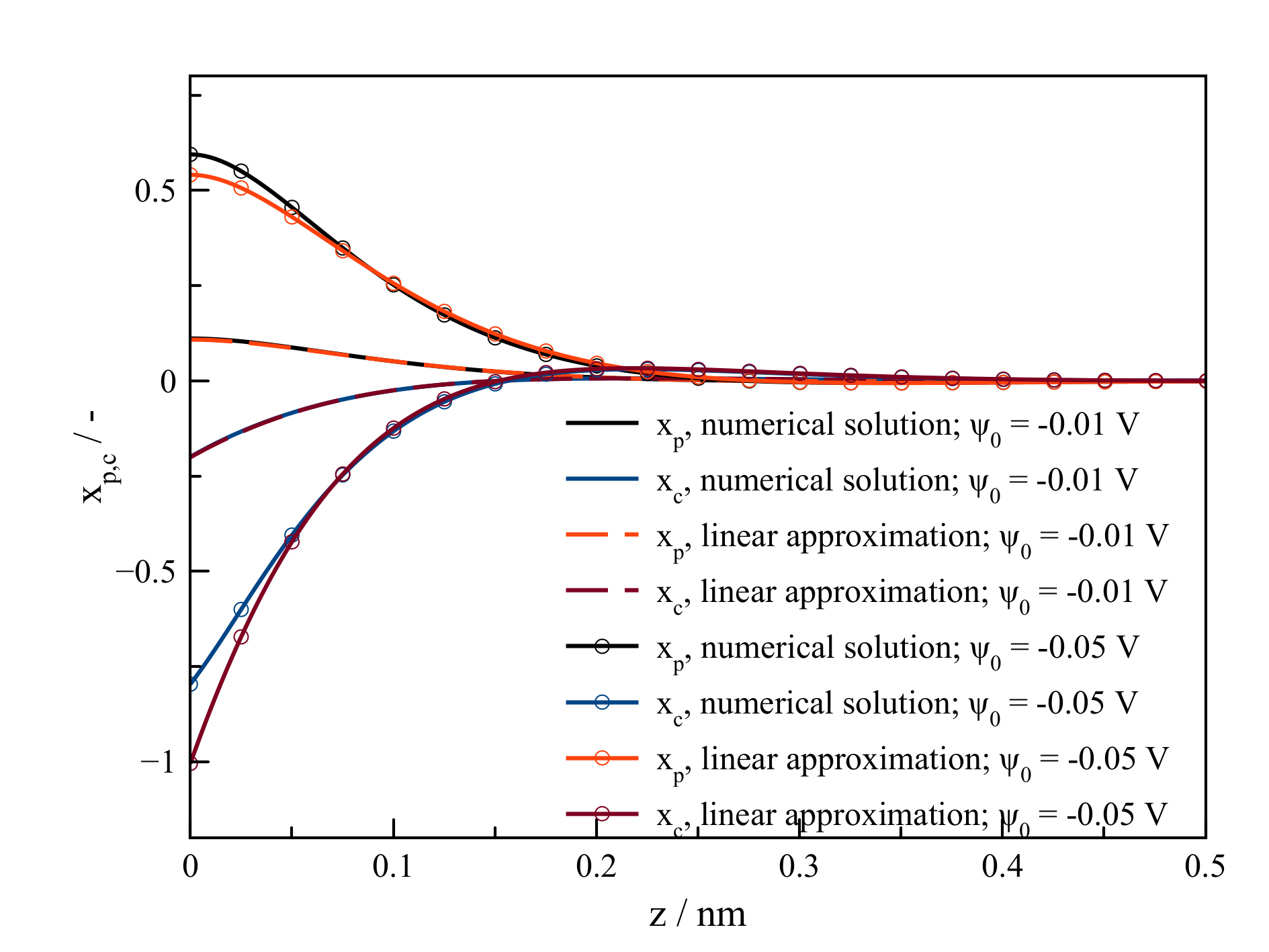}
\caption{Electrostatic potential (top) and concentration (bottom) profiles calculated within the linear analytical theory and nonlinear theory for negative potential drops. The data are shown for $\phi_0=0.4$, $b=v^{1/3}=0.5~nm$, $T=298~K$, $\varepsilon=5$, $q=1.6\times 10^{-19}~C$. The local concentrations are determined via $x_{p,c}(z)=(n_{p,c}(z)-n_0)/n_0$.}
\label{fig6} 
\end{figure}

Now we would like to discuss the electrostatic potential profiles and local ionic concentrations following from the numerical solution of the SCF equations (\ref{scf_eq_2}) and compare them with those predicted by the linear theory. As is seen in Fig. \ref{fig5}, at positive voltages the potential profile has a minimum which becomes more pronounced with a voltage increase. The same trends are observed for negative voltage case (see Fig. \ref{fig6}). Such potential profile behavior is determined by an interplay of electrostatics and conformation entropy of the macromolecules similar to the case of the electrostatic potential of a point-like charge immersed in a PIL (see eq. (\ref{psi})) or a polyelectrolyte solution \cite{Borue1988,muthukumar1996double}.  Note that the potential profiles predicted by the linear theory practically coincide with the numerical solutions at the voltages in the interval $|\psi_0|\leq 0.05~V$. At the voltages $|\psi_0|>0.05~V$ the linear theory already gives significant discrepancy for the local electrostatic potential values. However, the interval of potential drops, where the linear theory gives a satisfactory approximation for the concentration profiles is even narrower. Indeed, already at $\psi_0\approx \pm 0.05~V$ the local concentrations of ionic species predicted by the linear theory slightly deviate from the numerical solution near the electrode surface ($z=0$). The deviation increases at larger potential drops. We would like to note that eqs. (\ref{xp2}), (\ref{psi2}), and (\ref{xc2}), obtained within the linear theory, determine the asymptotic behavior of the potential profile and local ionic concentration at sufficiently large distances from the electrode at any applied voltages. It is interesting to note that the oscillation effect for the concentration profiles is very small, so that they are qualitatively the same as those predicted by Kornyshev's mean-field theory for regular ILs.

\begin{figure}[h]
\centering
\includegraphics[width=0.6\linewidth]{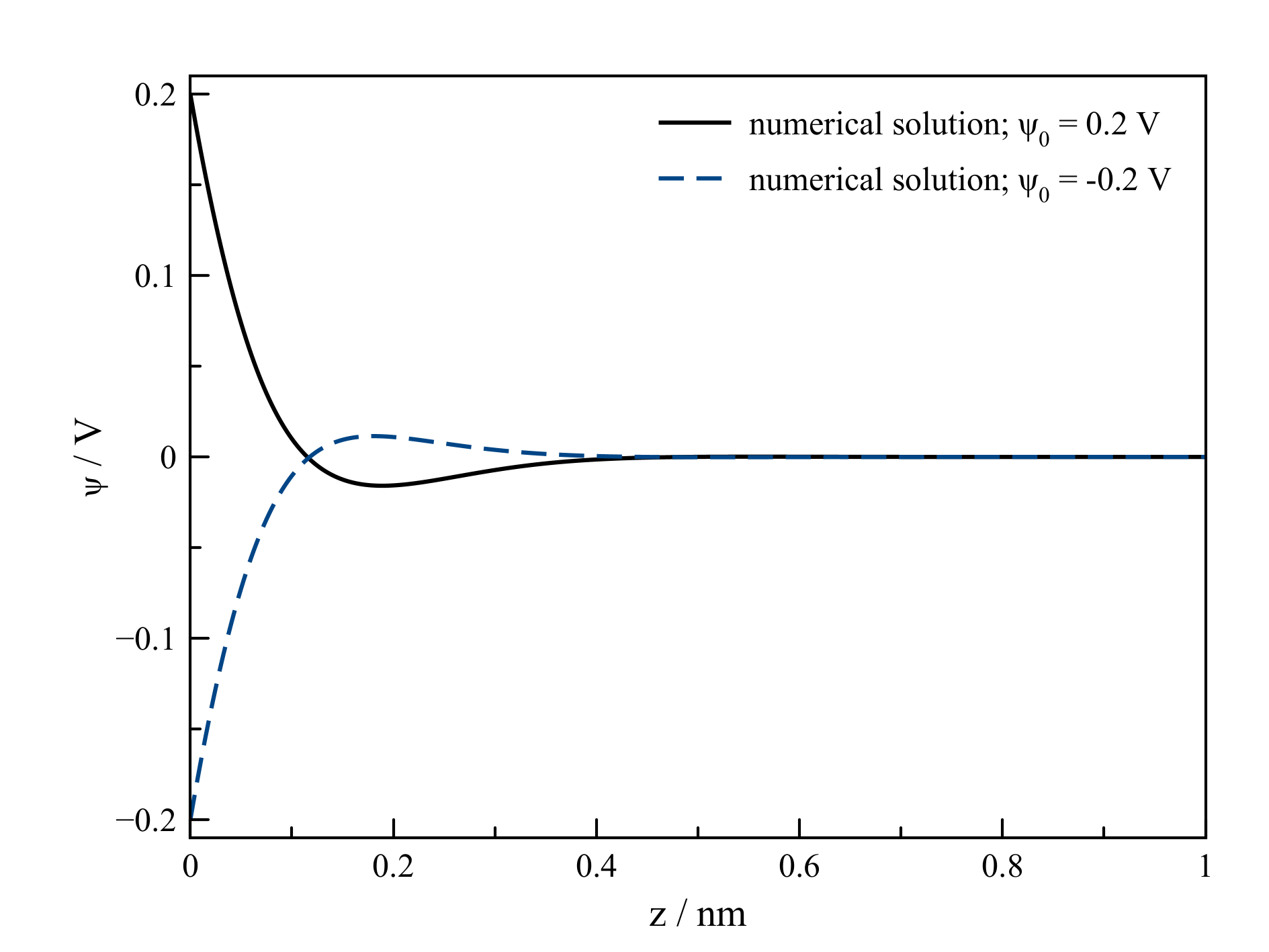}
\includegraphics[width=0.6\linewidth]{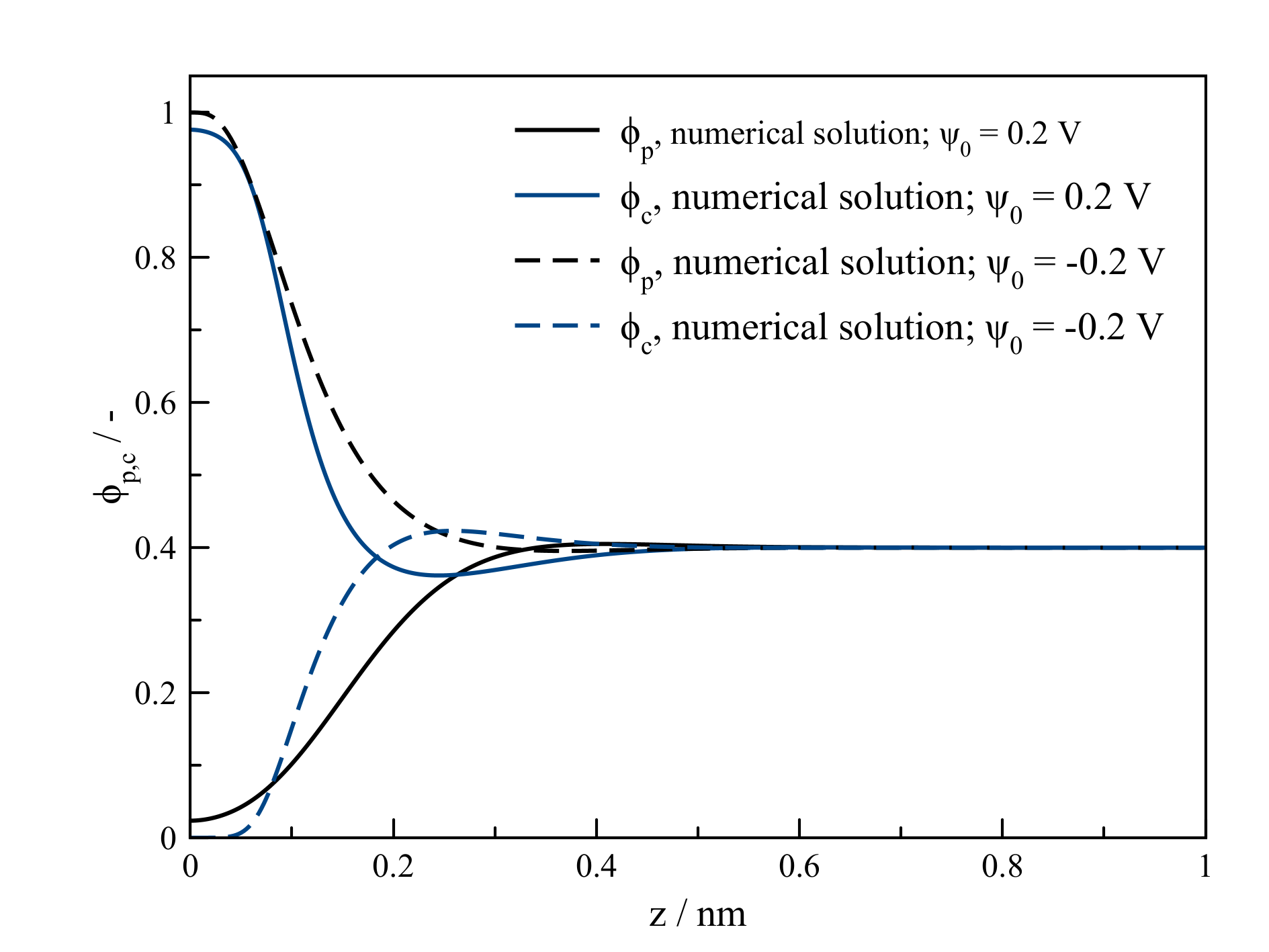}
\caption{Electrostatic potential (top) and volume fraction (bottom) profiles numerically calculated within the nonlinear theory for sufficiently high voltages $\psi_0=\pm 0.2~V$. The data are shown for $\phi_0=0.4$, $b=v^{1/3}=0.5~nm$, $T=298~K$, $\varepsilon=5$, $q=1.6\times 10^{-19}~C$.}
\label{fig7} 
\end{figure}

Finally, it is interesting to discuss the behavior of the potential and concentration profiles at a sufficiently large potential drop, at which the linear theory does not work. Fig. \ref{fig7} shows the electrostatic potential (top) and local concentrations (bottom) of species for the voltages $\psi_0=\pm 0.2~V$. As is seen, in contrast to the salt-free polyelectrolyte solution case, where the electrostatic potential or concentration decrease (increase) is accompanied by rather strong oscillations \cite{chatellier1996adsorption}, these quantities in the case of a pure PIL do not show pronounced oscillation behavior. Instead, there are quite pronounced maxima or minima at a certain distance from the electrode. It is also interesting to note that the potential profiles at $\psi_0=\pm 0.2~V$ are almost symmetric. Such symmetry can be explained by the fact that at a sufficiently high voltage the local volume fraction of the monomer units (for a negatively charged electrode) or counterions (for a positively charged electrode) near the electrode is close to unity so that it does not matter whether the particles are tied in a chain or freely move  -- the translation and conformation contributions to the entropy are equal to zero in both cases. We would like to underline once again that the almost symmetric form of the differential capacitance discussed above (see Fig. \ref{fig2}) has the same causes.

\begin{figure}[h]
\centering
\includegraphics[width=0.6\linewidth]{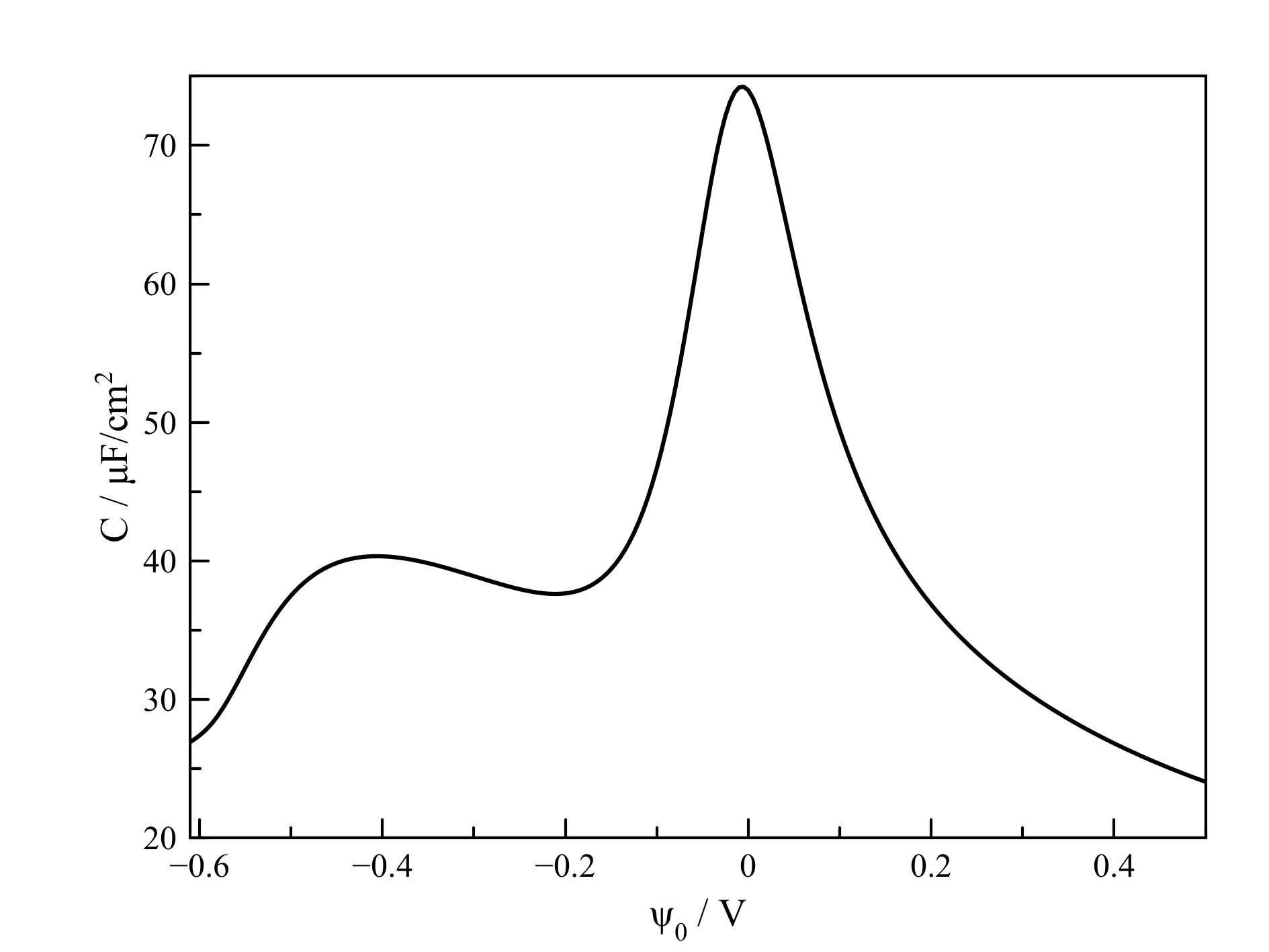}
\includegraphics[width=0.6\linewidth]{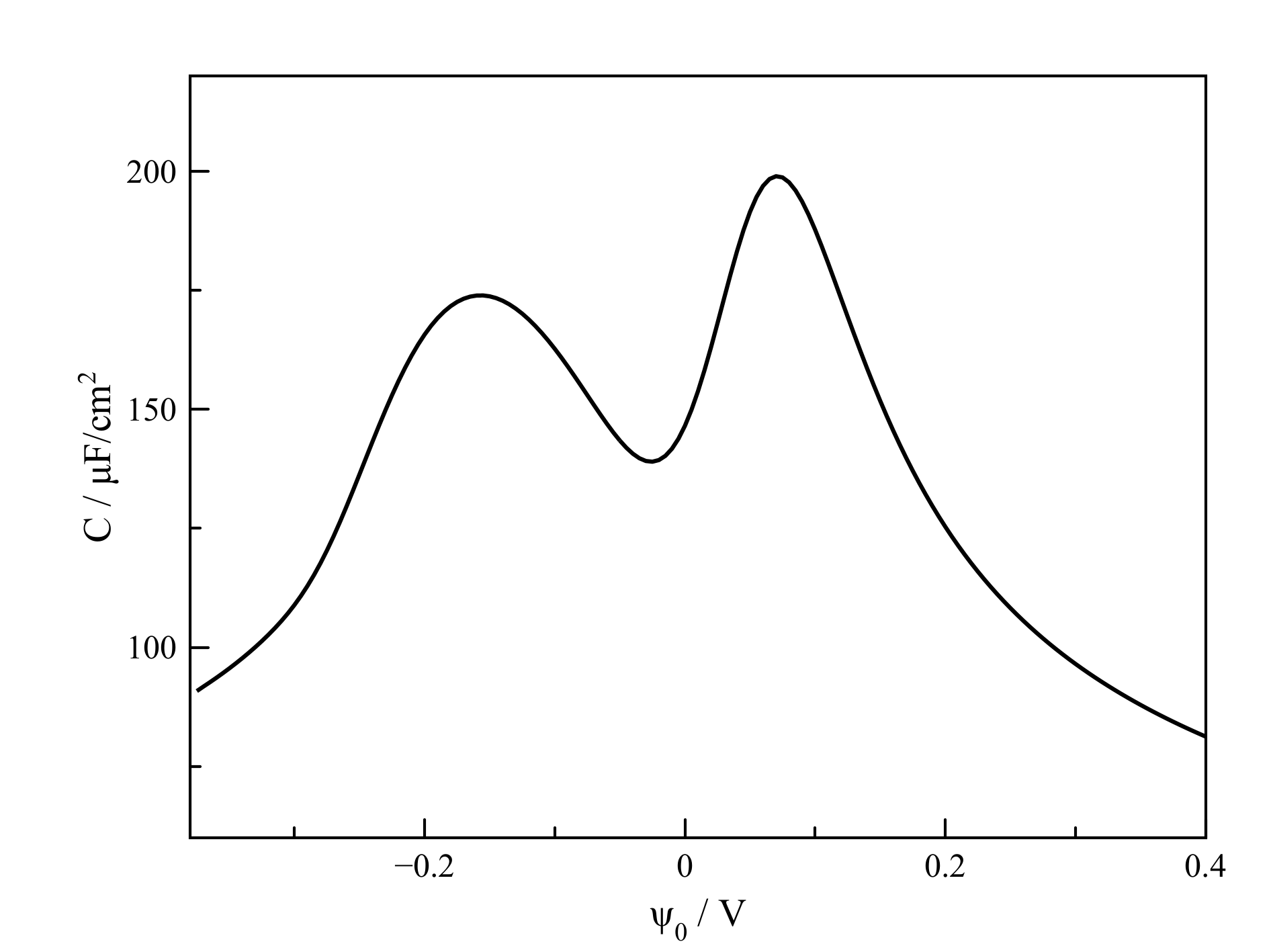}
\caption{{Differential capacitance profiles for a pure PIL (top) and the same PIL dissolved in an organic solvent (bottom) plotted for the hard wall regime. The physical parameters are the same as those taken for Figures \ref{fig1} and \ref{fig3}.}}
\label{fig8} 
\end{figure}

\begin{figure}[h]
\centering
\includegraphics[width=0.6\linewidth]{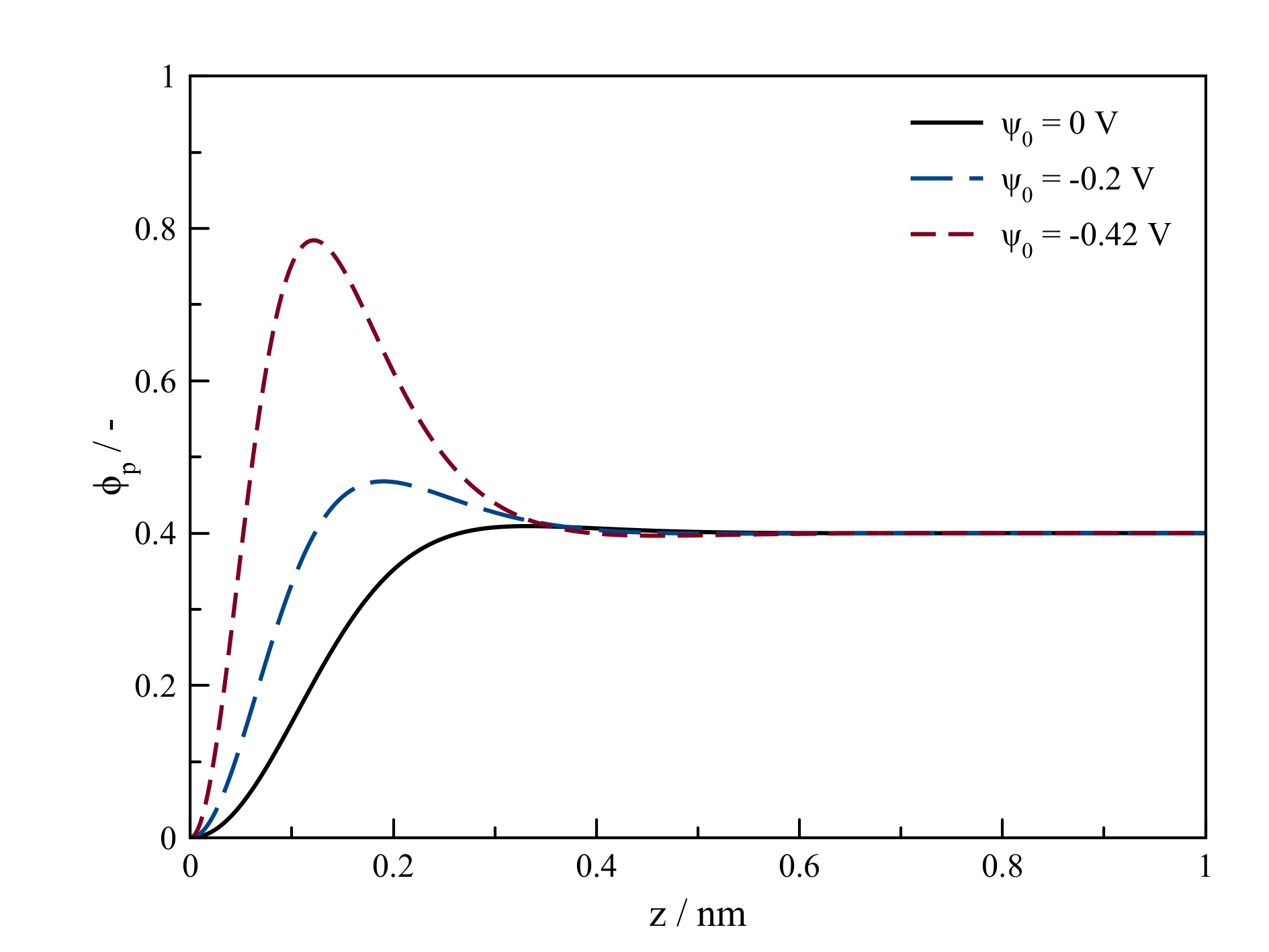}
\includegraphics[width=0.6\linewidth]{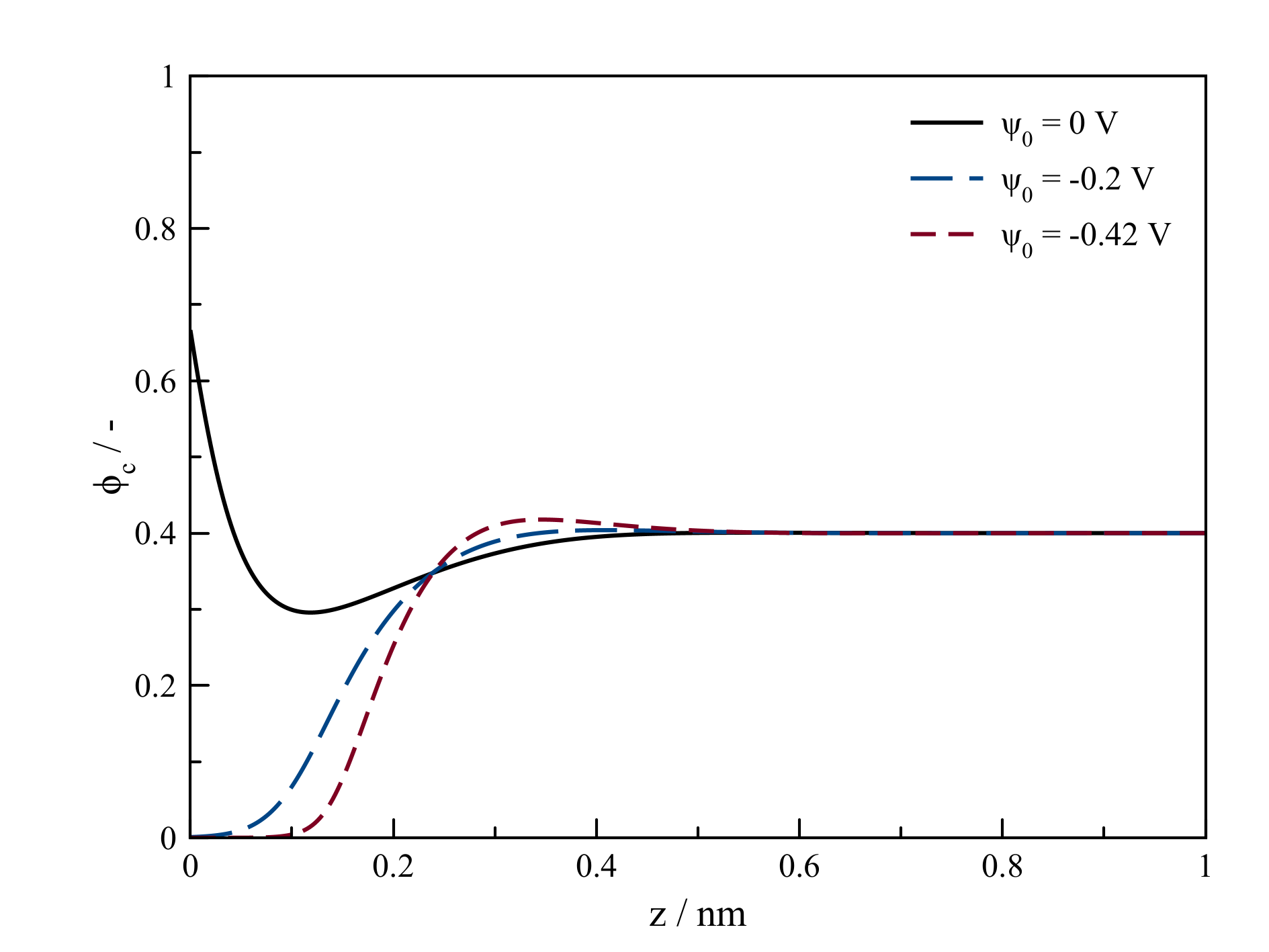}
\caption{{Local volume fraction profiles of monomer units (top) and counterions (bottom) for a pure PIL calculated in the hard wall regime. The data are shown for $\phi_0=0.4$, $b=v^{1/3}=0.5~nm$, $T=298~K$, $\varepsilon=5$, $q=1.6\times 10^{-19}~C$.}}
\label{fig9} 
\end{figure}

{\textbf{Hard wall case.} It is instructive to consider another limiting regime of the hard wall corresponding to the boundary condition $\phi_p(0)=0$. This boundary condition describes the occurrence of strong specific polymer-electrode repulsion at small distances. In this case, the linear theory discussed above does not work even at sufficiently small voltages due to the strong deviation of the local polymer volume fraction near the electrode from the bulk value, which is caused by the monomeric unit depletion near the electrode~\cite{netz2003neutral,Andelman2000}. As is seen in Fig.\ref{fig8}, the differential capacitance profiles are qualitatively different from those obtained in the indifferent electrode regime. In contrast to the case of a pure PIL on an indifferent electrode, where a slightly shifted bell-shaped differential capacitance profile is realized, in the hard wall regime the differential capacitance has a camel-shaped profile with two strongly asymmetric peaks (see Fig.\ref{fig8}(top)). One maximum, higher and narrower, is located in the vicinity of zero voltage with a slight shift to negative potential values, whereas the another one is in the region of sufficiently large negative voltages. For the case of a PIL solution, we obtain a standard camel-shaped differential capacitance profile with peaks relatively similar in shape. The nature of the wider and shorter peak in the case of the pure PIL can be understood by analysing the behavior of the local volume fractions of the monomeric units and counterions. As is seen in Fig. \ref{fig9}, at zero voltage, the monomer units are depleted near the electrode surface within the rather narrow layer. Counterions, on the contrary, are in excess in the same layer relative to the bulk. However, after the certain negative voltage ($\psi_0\approx -0.4~V$), the depletion layer becomes narrower and narrower and the maximum on the local polymer volume fraction profiles grows. In other words, at a sufficiently high negative voltage, the interplay between the depletion of the monomeric units and their Coulomb attraction to the electrode results in the formation of a polymer adsorption layer. At the same time, the counterions are expelled from the electrode surface due to the Coulomb repulsion and steric effect. The formation of a polymer adsorption layer is related to the maximum on the differential capacitance profile in the region of negative voltages. The case of a PIL dissolved in an organic solvent is similar to the one, observed for the indifferent surface boundary condition case except that the peak in the region of negative potential drops is a lot shorter. This is caused by the weaker polymer adsorption due to the polymer-surface short-range repulsion. Note that as in the indifferent electrode case considered above, at sufficiently large voltages, the same asymptotic behavior, $C\sim |\psi_0|^{-0.5}$, is observed.}

\section{Conclusions and prospects}
We have proposed a self-consistent field theory of the polymeric ionic liquid on a charged metal electrode. Taking into account the conformation entropy of polymerized cations within the Lifshitz theory and electrostatic and excluded volume interactions of ionic species within the asymmetric lattice gas model at the mean-field level of approximation, we have obtained a system of self-consistent field equations for the local electrostatic potential and average concentrations of monomeric units and counterions. In the linear approximation of the self-consistent field equations, for the cases of point-like charge and a flat infinite uniformly charged {indifferent electrode (without polymer-electrode specific interactions)} immersed in a polymeric ionic liquid, we have analyzed the behavior of local ionic concentrations and electrostatic potential. We have derived an analytical expression for the linear differential capacitance of the electric double layer. We have shown that in contrast to the linear capacitance predicted by Kornyshev's self-consistent field theory of regular ionic liquids on a charged electrode, linear capacitance for the case of polymeric ionic liquids strongly depends on the molecular parameters of polymerized cations: effective excluded volume of the monomeric unit and bond length of the polymeric cation. We have analyzed the differential capacitance behavior as a potential drop function. We have shown that {for the case of the indifferent surface} at moderate voltages slightly asymmetric bell-shaped differential capacitance profile is lower and wider than that predicted in Kornyshev's mean-field theory for monomeric ionic liquids. However, at a rather high voltage, the differential capacitance monotonically decreases in accordance with the same power law, $C\sim |\psi_0|^{-0.5}$, as for monomeric ionic liquids, predicted within Kornyshev's analytical self-consistent field theory and more sophisticated calculations based on the non-local density functional theory formalism \cite{wu2011classical}. For the case of a polymeric ionic liquid solution on a charged electrode {with the same boundary condition}, we have predicted a strongly asymmetric camel-shaped differential capacitance profile. We have obtained a rather high peak on the differential capacitance profile in the region of negative voltages and an unpronounced local maximum -- at positive potential drops. We have attributed the higher peak to the dramatic increase in the local volume fraction of the monomeric units on the electrode when moving towards higher negative voltage values, while the small peak -- to a rather smooth increase in the local volume fraction of the counterions when increasing the positive voltage. We have analyzed the electrostatic potential and concentration profiles for a pure polymeric ionic liquid demonstrating that in contrast to sufficiently dilute salt-free polyelectrolyte solutions \cite{chatellier1996adsorption}, these quantities have no pronounced oscillation behavior. For the case of the hard wall boundary condition, corresponding to the strong short-range repulsion of the monomeric units from the electrode surface, we obtained a noticeable contrast in the differential capacitance profiles for the pure PIL, expressed in the strong asymmetric camel-shaped differential capacitance profile, compared to the slightly shifted bell-shaped profile of the pure PIL in the case of the indifferent surface. Note that considering only the limiting regimes of the indifferent surface and the hard wall, we did not introduce into the theory polymer-electrode interaction energy or mentioned above extrapolation length which can have either sign. The influence of the polymer-electrode specific interactions (attractive or repulsive) on the differential capacitance profile will be systematically studied elsewhere.

{In conclusion, we would like to note that this study aims to elucidate the role of the chain connectivity (conformational entropy) of cations in the differential capacitance behavior for polymeric ionic liquids on a charged electrode at the simplest theoretical level. That is why we did not take into account the size asymmetry of the anions and cation monomeric segments, as in the monomeric ionic liquid case in ref. \cite{kornyshev2007double}. Note that the ionic size difference can be taken into account within the off-lattice equations of state of hard-spheres mixtures \cite{Maggs2016,boublik1970hard,mansoori1971equilibrium}. Moreover, like Kornyshev's theory, the present theory does not take into account the short-range specific interactions of the ions. The latter can be easily accounted for in the same manner as in refs. \cite{goodwin2017mean,budkov2018theory}. We also did not take explicitly into account the static polarizabilities and permanent dipole moments of the monomeric units, describing the electric polarization effect via the constant dielectric permittivity. The microscopic dielectric properties of the ionic species can be implemented into the self-consistent field theory as in refs. \cite{budkovJPCC2021,budkov2020two,budkov2015modified,budkov2016theory,budkov2018theory,abrashkin2007dipolar}. Finally, in this study, we did not consider the electrostatic correlations of ionic particles and nonlocal packing effects. These effects can be utilized within the more sophisticated nonlocal density functional theory \cite{Li2006,Li2006a} mentioned in the introduction. In particular, it is interesting to compare the differential capacitance profiles obtained within the framework of the nonlocal density functional theory with those obtained within the present mean-field theory. All of these issues could be the subject of the forthcoming publications.}

We believe that our findings can be of interesting for electrochemical applications, such as supercapacitors, batteries, fuel cells, catalysis, electrodeposition, {\sl etc.}

\section*{Acknowledgments}
The development of the self-consistent field theory presented in section 2 is supported by the grant of the President of the Russian Federation (project No. MD-341.2021.1.3). The results presented in sections 3 and 4 have been obtained with supporting of the Russian Science Foundation (Grant No. 21-11-00031).
\bibliography{name}

\begin{thebibliography}{65}%
\makeatletter
\providecommand \@ifxundefined [1]{%
 \@ifx{#1\undefined}
}%
\providecommand \@ifnum [1]{%
 \ifnum #1\expandafter \@firstoftwo
 \else \expandafter \@secondoftwo
 \fi
}%
\providecommand \@ifx [1]{%
 \ifx #1\expandafter \@firstoftwo
 \else \expandafter \@secondoftwo
 \fi
}%
\providecommand \natexlab [1]{#1}%
\providecommand \enquote  [1]{``#1''}%
\providecommand \bibnamefont  [1]{#1}%
\providecommand \bibfnamefont [1]{#1}%
\providecommand \citenamefont [1]{#1}%
\providecommand \href@noop [0]{\@secondoftwo}%
\providecommand \href [0]{\begingroup \@sanitize@url \@href}%
\providecommand \@href[1]{\@@startlink{#1}\@@href}%
\providecommand \@@href[1]{\endgroup#1\@@endlink}%
\providecommand \@sanitize@url [0]{\catcode `\\12\catcode `\$12\catcode
  `\&12\catcode `\#12\catcode `\^12\catcode `\_12\catcode `\%12\relax}%
\providecommand \@@startlink[1]{}%
\providecommand \@@endlink[0]{}%
\providecommand \url  [0]{\begingroup\@sanitize@url \@url }%
\providecommand \@url [1]{\endgroup\@href {#1}{\urlprefix }}%
\providecommand \urlprefix  [0]{URL }%
\providecommand \Eprint [0]{\href }%
\providecommand \doibase [0]{http://dx.doi.org/}%
\providecommand \selectlanguage [0]{\@gobble}%
\providecommand \bibinfo  [0]{\@secondoftwo}%
\providecommand \bibfield  [0]{\@secondoftwo}%
\providecommand \translation [1]{[#1]}%
\providecommand \BibitemOpen [0]{}%
\providecommand \bibitemStop [0]{}%
\providecommand \bibitemNoStop [0]{.\EOS\space}%
\providecommand \EOS [0]{\spacefactor3000\relax}%
\providecommand \BibitemShut  [1]{\csname bibitem#1\endcsname}%
\let\auto@bib@innerbib\@empty
\bibitem [{\citenamefont {Fedorov}\ and\ \citenamefont
  {Kornyshev}(2014)}]{fedorov2014ionic}%
  \BibitemOpen
  \bibfield  {author} {\bibinfo {author} {\bibfnamefont {M.~V.}\ \bibnamefont
  {Fedorov}}\ and\ \bibinfo {author} {\bibfnamefont {A.~A.}\ \bibnamefont
  {Kornyshev}},\ }\bibfield  {title} {\enquote {\bibinfo {title} {Ionic liquids
  at electrified interfaces},}\ }\href@noop {} {\bibfield  {journal} {\bibinfo
  {journal} {Chemical reviews}\ }\textbf {\bibinfo {volume} {114}},\ \bibinfo
  {pages} {2978--3036} (\bibinfo {year} {2014})}\BibitemShut {NoStop}%
\bibitem [{\citenamefont {Ohno}\ and\ \citenamefont {Ito}(1998)}]{ohno1998}%
  \BibitemOpen
  \bibfield  {author} {\bibinfo {author} {\bibfnamefont {H.}~\bibnamefont
  {Ohno}}\ and\ \bibinfo {author} {\bibfnamefont {K.}~\bibnamefont {Ito}},\
  }\bibfield  {title} {\enquote {\bibinfo {title} {{Room-temperature molten
  salt polymers as a matrix for fast ion conduction}},}\ }\bibfield
  {booktitle} {\emph {\bibinfo {booktitle} {Chemistry Letters}},\ }\href
  {\doibase 10.1246/cl.1998.751} {\ ,\ \bibinfo {pages} {751--752} (\bibinfo
  {year} {1998})}\BibitemShut {NoStop}%
\bibitem [{\citenamefont {Ohno}, \citenamefont {Yoshizawa},\ and\ \citenamefont
  {Ogihara}(2004)}]{ohno2004}%
  \BibitemOpen
  \bibfield  {author} {\bibinfo {author} {\bibfnamefont {H.}~\bibnamefont
  {Ohno}}, \bibinfo {author} {\bibfnamefont {M.}~\bibnamefont {Yoshizawa}}, \
  and\ \bibinfo {author} {\bibfnamefont {W.}~\bibnamefont {Ogihara}},\
  }\bibfield  {title} {\enquote {\bibinfo {title} {{Development of new class of
  ion conductive polymers based on ionic liquids}},}\ }\href {\doibase
  10.1016/j.electacta.2004.01.091} {\bibfield  {journal} {\bibinfo  {journal}
  {Electrochimica Acta}\ }\textbf {\bibinfo {volume} {50}},\ \bibinfo {pages}
  {255--261} (\bibinfo {year} {2004})}\BibitemShut {NoStop}%
\bibitem [{\citenamefont {Ohno}(2007)}]{ohno2007}%
  \BibitemOpen
  \bibfield  {author} {\bibinfo {author} {\bibfnamefont {H.}~\bibnamefont
  {Ohno}},\ }\bibfield  {title} {\enquote {\bibinfo {title} {{Design of ion
  conductive polymers based on ionic liquids}},}\ }\href {\doibase
  10.1002/masy.200750435} {\bibfield  {journal} {\bibinfo  {journal}
  {Macromolecular Symposia}\ }\textbf {\bibinfo {volume} {249-250}},\ \bibinfo
  {pages} {551--556} (\bibinfo {year} {2007})}\BibitemShut {NoStop}%
\bibitem [{\citenamefont {Yuan}\ and\ \citenamefont
  {Antonietti}(2011)}]{Yuan2011}%
  \BibitemOpen
  \bibfield  {author} {\bibinfo {author} {\bibfnamefont {J.}~\bibnamefont
  {Yuan}}\ and\ \bibinfo {author} {\bibfnamefont {M.}~\bibnamefont
  {Antonietti}},\ }\bibfield  {title} {\enquote {\bibinfo {title} {{Poly(ionic
  liquid)s: Polymers expanding classical property profiles}},}\ }\href
  {\doibase 10.1016/J.POLYMER.2011.01.043} {\bibfield  {journal} {\bibinfo
  {journal} {Polymer}\ }\textbf {\bibinfo {volume} {52}},\ \bibinfo {pages}
  {1469--1482} (\bibinfo {year} {2011})}\BibitemShut {NoStop}%
\bibitem [{\citenamefont {Mecerreyes}(2011)}]{Mecerreyes2011}%
  \BibitemOpen
  \bibfield  {author} {\bibinfo {author} {\bibfnamefont {D.}~\bibnamefont
  {Mecerreyes}},\ }\bibfield  {title} {\enquote {\bibinfo {title} {{Polymeric
  ionic liquids: Broadening the properties and applications of
  polyelectrolytes}},}\ }\href {\doibase 10.1016/J.PROGPOLYMSCI.2011.05.007}
  {\bibfield  {journal} {\bibinfo  {journal} {Progress in Polymer Science}\
  }\textbf {\bibinfo {volume} {36}},\ \bibinfo {pages} {1629--1648} (\bibinfo
  {year} {2011})}\BibitemShut {NoStop}%
\bibitem [{\citenamefont {Eftekhari}\ and\ \citenamefont
  {Saito}(2017)}]{Eftekhari2017}%
  \BibitemOpen
  \bibfield  {author} {\bibinfo {author} {\bibfnamefont {A.}~\bibnamefont
  {Eftekhari}}\ and\ \bibinfo {author} {\bibfnamefont {T.}~\bibnamefont
  {Saito}},\ }\bibfield  {title} {\enquote {\bibinfo {title} {{Synthesis and
  properties of polymerized ionic liquids}},}\ }\href {\doibase
  10.1016/j.eurpolymj.2017.03.033} {\bibfield  {journal} {\bibinfo  {journal}
  {European Polymer Journal}\ }\textbf {\bibinfo {volume} {90}},\ \bibinfo
  {pages} {245--272} (\bibinfo {year} {2017})}\BibitemShut {NoStop}%
\bibitem [{\citenamefont {Fadeeva}\ \emph {et~al.}(2021)\citenamefont
  {Fadeeva}, \citenamefont {Shmukler}, \citenamefont {Gruzdev},\ and\
  \citenamefont {Safonova}}]{fadeeva2021imidazolium}%
  \BibitemOpen
  \bibfield  {author} {\bibinfo {author} {\bibfnamefont {Y.~A.}\ \bibnamefont
  {Fadeeva}}, \bibinfo {author} {\bibfnamefont {L.~E.}\ \bibnamefont
  {Shmukler}}, \bibinfo {author} {\bibfnamefont {M.~S.}\ \bibnamefont
  {Gruzdev}}, \ and\ \bibinfo {author} {\bibfnamefont {L.~P.}\ \bibnamefont
  {Safonova}},\ }\bibfield  {title} {\enquote {\bibinfo {title} {Imidazolium
  zwitterion-based protic ionic liquids: from monomers to polymer membranes},}\
  }\href@noop {} {\bibfield  {journal} {\bibinfo  {journal} {Polymer
  International}\ }\textbf {\bibinfo {volume} {70}},\ \bibinfo {pages}
  {1582--1589} (\bibinfo {year} {2021})}\BibitemShut {NoStop}%
\bibitem [{\citenamefont {Chen}\ \emph {et~al.}(2018)\citenamefont {Chen},
  \citenamefont {Frenzel}, \citenamefont {Cui}, \citenamefont {Gao},
  \citenamefont {Campanella}, \citenamefont {Funtan}, \citenamefont {Kremer},
  \citenamefont {Parkin},\ and\ \citenamefont {Binder}}]{chen2018}%
  \BibitemOpen
  \bibfield  {author} {\bibinfo {author} {\bibfnamefont {S.}~\bibnamefont
  {Chen}}, \bibinfo {author} {\bibfnamefont {F.}~\bibnamefont {Frenzel}},
  \bibinfo {author} {\bibfnamefont {B.}~\bibnamefont {Cui}}, \bibinfo {author}
  {\bibfnamefont {F.}~\bibnamefont {Gao}}, \bibinfo {author} {\bibfnamefont
  {A.}~\bibnamefont {Campanella}}, \bibinfo {author} {\bibfnamefont
  {A.}~\bibnamefont {Funtan}}, \bibinfo {author} {\bibfnamefont
  {F.}~\bibnamefont {Kremer}}, \bibinfo {author} {\bibfnamefont {S.~S.}\
  \bibnamefont {Parkin}}, \ and\ \bibinfo {author} {\bibfnamefont {W.~H.}\
  \bibnamefont {Binder}},\ }\bibfield  {title} {\enquote {\bibinfo {title}
  {{Gating effects of conductive polymeric ionic liquids}},}\ }\href {\doibase
  10.1039/c8tc01936c} {\bibfield  {journal} {\bibinfo  {journal} {Journal of
  Materials Chemistry C}\ }\textbf {\bibinfo {volume} {6}},\ \bibinfo {pages}
  {8242--8250} (\bibinfo {year} {2018})}\BibitemShut {NoStop}%
\bibitem [{\citenamefont {Peltekoff}\ \emph {et~al.}(2019)\citenamefont
  {Peltekoff}, \citenamefont {Hiller}, \citenamefont {Lopinski}, \citenamefont
  {Melville},\ and\ \citenamefont {Lessard}}]{peltekoff2019}%
  \BibitemOpen
  \bibfield  {author} {\bibinfo {author} {\bibfnamefont {A.~J.}\ \bibnamefont
  {Peltekoff}}, \bibinfo {author} {\bibfnamefont {V.~E.}\ \bibnamefont
  {Hiller}}, \bibinfo {author} {\bibfnamefont {G.~P.}\ \bibnamefont
  {Lopinski}}, \bibinfo {author} {\bibfnamefont {O.~A.}\ \bibnamefont
  {Melville}}, \ and\ \bibinfo {author} {\bibfnamefont {B.~H.}\ \bibnamefont
  {Lessard}},\ }\bibfield  {title} {\enquote {\bibinfo {title} {{Unipolar
  Polymerized Ionic Liquid Copolymers as High-Capacitance Electrolyte Gates for
  n-Type Transistors}},}\ }\href {\doibase 10.1021/acsapm.9b00959} {\bibfield
  {journal} {\bibinfo  {journal} {ACS Applied Polymer Materials}\ }\textbf
  {\bibinfo {volume} {1}},\ \bibinfo {pages} {3210--3221} (\bibinfo {year}
  {2019})}\BibitemShut {NoStop}%
\bibitem [{\citenamefont {Gupta}, \citenamefont {Liang},\ and\ \citenamefont
  {Hu}(2019)}]{gupta2019}%
  \BibitemOpen
  \bibfield  {author} {\bibinfo {author} {\bibfnamefont {N.}~\bibnamefont
  {Gupta}}, \bibinfo {author} {\bibfnamefont {Y.~N.}\ \bibnamefont {Liang}}, \
  and\ \bibinfo {author} {\bibfnamefont {X.}~\bibnamefont {Hu}},\ }\bibfield
  {title} {\enquote {\bibinfo {title} {{Thermally responsive ionic liquids and
  polymeric ionic liquids: emerging trends and possibilities}},}\ }\href
  {\doibase 10.1016/j.coche.2019.07.005} {\bibfield  {journal} {\bibinfo
  {journal} {Current Opinion in Chemical Engineering}\ }\textbf {\bibinfo
  {volume} {25}},\ \bibinfo {pages} {43--50} (\bibinfo {year}
  {2019})}\BibitemShut {NoStop}%
\bibitem [{\citenamefont {{Yuki Kohno}}\ \emph {et~al.}(2015)\citenamefont
  {{Yuki Kohno}}, \citenamefont {{Shohei Saita}}, \citenamefont {{Yongjun
  Men}}, \citenamefont {{Jiayin Yuan}},\ and\ \citenamefont {{Hiroyuki
  Ohno}}}]{YukiKohno2015}%
  \BibitemOpen
  \bibfield  {author} {\bibinfo {author} {\bibnamefont {{Yuki Kohno}}},
  \bibinfo {author} {\bibnamefont {{Shohei Saita}}}, \bibinfo {author}
  {\bibnamefont {{Yongjun Men}}}, \bibinfo {author} {\bibnamefont {{Jiayin
  Yuan}}}, \ and\ \bibinfo {author} {\bibnamefont {{Hiroyuki Ohno}}},\
  }\bibfield  {title} {\enquote {\bibinfo {title} {{Thermoresponsive
  polyelectrolytes derived from ionic liquids}},}\ }\href {\doibase
  10.1039/C4PY01665C} {\bibfield  {journal} {\bibinfo  {journal} {Polymer
  Chemistry}\ }\textbf {\bibinfo {volume} {6}},\ \bibinfo {pages} {2163--2178}
  (\bibinfo {year} {2015})}\BibitemShut {NoStop}%
\bibitem [{\citenamefont {Tang}\ \emph {et~al.}(2014)\citenamefont {Tang},
  \citenamefont {Liu}, \citenamefont {Guo}, \citenamefont {Liu},\ and\
  \citenamefont {Jiang}}]{Tang2014}%
  \BibitemOpen
  \bibfield  {author} {\bibinfo {author} {\bibfnamefont {S.}~\bibnamefont
  {Tang}}, \bibinfo {author} {\bibfnamefont {S.}~\bibnamefont {Liu}}, \bibinfo
  {author} {\bibfnamefont {Y.}~\bibnamefont {Guo}}, \bibinfo {author}
  {\bibfnamefont {X.}~\bibnamefont {Liu}}, \ and\ \bibinfo {author}
  {\bibfnamefont {S.}~\bibnamefont {Jiang}},\ }\bibfield  {title} {\enquote
  {\bibinfo {title} {{Recent advances of ionic liquids and polymeric ionic
  liquids in capillary electrophoresis and capillary electrochromatography}},}\
  }\href {\doibase 10.1016/j.chroma.2014.04.037} {\bibfield  {journal}
  {\bibinfo  {journal} {Journal of Chromatography A}\ }\textbf {\bibinfo
  {volume} {1357}},\ \bibinfo {pages} {147--157} (\bibinfo {year}
  {2014})}\BibitemShut {NoStop}%
\bibitem [{\citenamefont {Gao}\ \emph {et~al.}(2021)\citenamefont {Gao},
  \citenamefont {Itliong}, \citenamefont {Kumar}, \citenamefont {Hjorth},\ and\
  \citenamefont {Nakamura}}]{gao2021polarization}%
  \BibitemOpen
  \bibfield  {author} {\bibinfo {author} {\bibfnamefont {T.}~\bibnamefont
  {Gao}}, \bibinfo {author} {\bibfnamefont {J.}~\bibnamefont {Itliong}},
  \bibinfo {author} {\bibfnamefont {S.~P.}\ \bibnamefont {Kumar}}, \bibinfo
  {author} {\bibfnamefont {Z.}~\bibnamefont {Hjorth}}, \ and\ \bibinfo {author}
  {\bibfnamefont {I.}~\bibnamefont {Nakamura}},\ }\bibfield  {title} {\enquote
  {\bibinfo {title} {Polarization of ionic liquid and polymer and its
  implications for polymerized ionic liquids: An overview towards a new theory
  and simulation},}\ }\href@noop {} {\bibfield  {journal} {\bibinfo  {journal}
  {Journal of Polymer Science}\ } (\bibinfo {year} {2021})}\BibitemShut
  {NoStop}%
\bibitem [{\citenamefont {Dobrynin}\ and\ \citenamefont
  {Rubinstein}(2005)}]{Dobrynin2005}%
  \BibitemOpen
  \bibfield  {author} {\bibinfo {author} {\bibfnamefont {A.~V.}\ \bibnamefont
  {Dobrynin}}\ and\ \bibinfo {author} {\bibfnamefont {M.}~\bibnamefont
  {Rubinstein}},\ }\bibfield  {title} {\enquote {\bibinfo {title} {{Theory of
  polyelectrolytes in solutions and at surfaces}},}\ }\href {\doibase
  10.1016/j.progpolymsci.2005.07.006} {\bibfield  {journal} {\bibinfo
  {journal} {Progress in Polymer Science (Oxford)}\ }\textbf {\bibinfo {volume}
  {30}},\ \bibinfo {pages} {1049--1118} (\bibinfo {year} {2005})}\BibitemShut
  {NoStop}%
\bibitem [{\citenamefont {Netz}\ and\ \citenamefont
  {Andelman}(2003)}]{netz2003neutral}%
  \BibitemOpen
  \bibfield  {author} {\bibinfo {author} {\bibfnamefont {R.~R.}\ \bibnamefont
  {Netz}}\ and\ \bibinfo {author} {\bibfnamefont {D.}~\bibnamefont
  {Andelman}},\ }\bibfield  {title} {\enquote {\bibinfo {title} {{Neutral and
  charged polymers at interfaces}},}\ }\href {\doibase
  10.1016/S0370-1573(03)00118-2} {\bibfield  {journal} {\bibinfo  {journal}
  {Physics Reports}\ }\textbf {\bibinfo {volume} {380}},\ \bibinfo {pages}
  {1--95} (\bibinfo {year} {2003})},\ \Eprint {http://arxiv.org/abs/0203364}
  {arXiv:0203364 [cond-mat]} \BibitemShut {NoStop}%
\bibitem [{\citenamefont {Andelman}\ and\ \citenamefont
  {Joanny}(2000{\natexlab{a}})}]{andelman2000polyelectrolyte}%
  \BibitemOpen
  \bibfield  {author} {\bibinfo {author} {\bibfnamefont {D.}~\bibnamefont
  {Andelman}}\ and\ \bibinfo {author} {\bibfnamefont {J.-F.}\ \bibnamefont
  {Joanny}},\ }\bibfield  {title} {\enquote {\bibinfo {title} {Polyelectrolyte
  adsorption},}\ }\href@noop {} {\bibfield  {journal} {\bibinfo  {journal}
  {Comptes Rendus de l'Acad{\'e}mie des Sciences-Series IV-Physics}\ }\textbf
  {\bibinfo {volume} {1}},\ \bibinfo {pages} {1153--1162} (\bibinfo {year}
  {2000}{\natexlab{a}})}\BibitemShut {NoStop}%
\bibitem [{\citenamefont {{Van Der Schee}}\ and\ \citenamefont
  {Lyklema}(1984)}]{vanderschee1984}%
  \BibitemOpen
  \bibfield  {author} {\bibinfo {author} {\bibfnamefont {H.~A.}\ \bibnamefont
  {{Van Der Schee}}}\ and\ \bibinfo {author} {\bibfnamefont {J.}~\bibnamefont
  {Lyklema}},\ }\bibfield  {title} {\enquote {\bibinfo {title} {{A lattice
  theory of polyelectrolyte adsorption}},}\ }\href {\doibase
  10.1021/j150670a031} {\bibfield  {journal} {\bibinfo  {journal} {Journal of
  Physical Chemistry}\ }\textbf {\bibinfo {volume} {88}},\ \bibinfo {pages}
  {6661--6667} (\bibinfo {year} {1984})}\BibitemShut {NoStop}%
\bibitem [{\citenamefont {Papenhuijzen}, \citenamefont {{Van Der Schee}},\ and\
  \citenamefont {Fleer}(1985)}]{Papenhuijzen1985}%
  \BibitemOpen
  \bibfield  {author} {\bibinfo {author} {\bibfnamefont {J.}~\bibnamefont
  {Papenhuijzen}}, \bibinfo {author} {\bibfnamefont {H.~A.}\ \bibnamefont {{Van
  Der Schee}}}, \ and\ \bibinfo {author} {\bibfnamefont {G.~J.}\ \bibnamefont
  {Fleer}},\ }\bibfield  {title} {\enquote {\bibinfo {title} {{Polyelectrolyte
  adsorption. I. A new lattice theory}},}\ }\href {\doibase
  10.1016/0021-9797(85)90061-X} {\bibfield  {journal} {\bibinfo  {journal}
  {Journal of Colloid And Interface Science}\ }\textbf {\bibinfo {volume}
  {104}},\ \bibinfo {pages} {540--552} (\bibinfo {year} {1985})}\BibitemShut
  {NoStop}%
\bibitem [{\citenamefont {Evers}\ \emph {et~al.}(1986)\citenamefont {Evers},
  \citenamefont {Fleer}, \citenamefont {Scheutjens},\ and\ \citenamefont
  {Lyklema}}]{evers1986}%
  \BibitemOpen
  \bibfield  {author} {\bibinfo {author} {\bibfnamefont {O.~A.}\ \bibnamefont
  {Evers}}, \bibinfo {author} {\bibfnamefont {G.~J.}\ \bibnamefont {Fleer}},
  \bibinfo {author} {\bibfnamefont {J.~M.}\ \bibnamefont {Scheutjens}}, \ and\
  \bibinfo {author} {\bibfnamefont {J.}~\bibnamefont {Lyklema}},\ }\bibfield
  {title} {\enquote {\bibinfo {title} {{Adsorption of weak polyelectrolytes
  from aqueous solution}},}\ }\href {\doibase 10.1016/0021-9797(86)90047-0}
  {\bibfield  {journal} {\bibinfo  {journal} {Journal of Colloid And Interface
  Science}\ }\textbf {\bibinfo {volume} {111}},\ \bibinfo {pages} {446--454}
  (\bibinfo {year} {1986})}\BibitemShut {NoStop}%
\bibitem [{\citenamefont {Roe}(1974)}]{roe1974}%
  \BibitemOpen
  \bibfield  {author} {\bibinfo {author} {\bibfnamefont {R.~J.}\ \bibnamefont
  {Roe}},\ }\bibfield  {title} {\enquote {\bibinfo {title} {{Multilayer theory
  of adsorption from a polymer solution}},}\ }\href {\doibase
  10.1063/1.1680888} {\bibfield  {journal} {\bibinfo  {journal} {The Journal of
  Chemical Physics}\ }\textbf {\bibinfo {volume} {60}},\ \bibinfo {pages}
  {4192--4207} (\bibinfo {year} {1974})}\BibitemShut {NoStop}%
\bibitem [{\citenamefont {Scheutjens}\ and\ \citenamefont
  {Fleer}(1979)}]{scheutjens1979statistical}%
  \BibitemOpen
  \bibfield  {author} {\bibinfo {author} {\bibfnamefont {J.}~\bibnamefont
  {Scheutjens}}\ and\ \bibinfo {author} {\bibfnamefont {G.}~\bibnamefont
  {Fleer}},\ }\bibfield  {title} {\enquote {\bibinfo {title} {Statistical
  theory of the adsorption of interacting chain molecules. 1. partition
  function, segment density distribution, and adsorption isotherms},}\
  }\href@noop {} {\bibfield  {journal} {\bibinfo  {journal} {Journal of
  Physical Chemistry}\ }\textbf {\bibinfo {volume} {83}},\ \bibinfo {pages}
  {1619--1635} (\bibinfo {year} {1979})}\BibitemShut {NoStop}%
\bibitem [{\citenamefont {Scheutjens}\ and\ \citenamefont
  {Fleer}(1980)}]{Scheutjens1980}%
  \BibitemOpen
  \bibfield  {author} {\bibinfo {author} {\bibfnamefont {J.~M.}\ \bibnamefont
  {Scheutjens}}\ and\ \bibinfo {author} {\bibfnamefont {G.~J.}\ \bibnamefont
  {Fleer}},\ }\bibfield  {title} {\enquote {\bibinfo {title} {{Statistical
  theory of the adsorption of interacting chain molecules. 2. Train, loop, and
  tail size distribution}},}\ }\href {\doibase 10.1021/j100439a011} {\bibfield
  {journal} {\bibinfo  {journal} {Journal of Physical Chemistry}\ }\textbf
  {\bibinfo {volume} {84}},\ \bibinfo {pages} {178--190} (\bibinfo {year}
  {1980})}\BibitemShut {NoStop}%
\bibitem [{\citenamefont {Borisov}, \citenamefont {Zhulina},\ and\
  \citenamefont {Birshtein}(1994)}]{borisov1994polyelectrolyte}%
  \BibitemOpen
  \bibfield  {author} {\bibinfo {author} {\bibfnamefont {O.}~\bibnamefont
  {Borisov}}, \bibinfo {author} {\bibfnamefont {E.}~\bibnamefont {Zhulina}}, \
  and\ \bibinfo {author} {\bibfnamefont {T.}~\bibnamefont {Birshtein}},\
  }\bibfield  {title} {\enquote {\bibinfo {title} {Polyelectrolyte molecule
  conformation near a charged surface},}\ }\href@noop {} {\bibfield  {journal}
  {\bibinfo  {journal} {Journal de Physique II}\ }\textbf {\bibinfo {volume}
  {4}},\ \bibinfo {pages} {913--929} (\bibinfo {year} {1994})}\BibitemShut
  {NoStop}%
\bibitem [{\citenamefont {Friedsam}, \citenamefont {Gaub},\ and\ \citenamefont
  {Netz}(2005)}]{friedsam2005adsorption}%
  \BibitemOpen
  \bibfield  {author} {\bibinfo {author} {\bibfnamefont {C.}~\bibnamefont
  {Friedsam}}, \bibinfo {author} {\bibfnamefont {H.}~\bibnamefont {Gaub}}, \
  and\ \bibinfo {author} {\bibfnamefont {R.}~\bibnamefont {Netz}},\ }\bibfield
  {title} {\enquote {\bibinfo {title} {Adsorption energies of single charged
  polymers},}\ }\href@noop {} {\bibfield  {journal} {\bibinfo  {journal} {EPL
  (Europhysics Letters)}\ }\textbf {\bibinfo {volume} {72}},\ \bibinfo {pages}
  {844} (\bibinfo {year} {2005})}\BibitemShut {NoStop}%
\bibitem [{\citenamefont {Varoqui}, \citenamefont {Johner},\ and\ \citenamefont
  {Elaissari}(1991)}]{varoqui1991conformation}%
  \BibitemOpen
  \bibfield  {author} {\bibinfo {author} {\bibfnamefont {R.}~\bibnamefont
  {Varoqui}}, \bibinfo {author} {\bibfnamefont {A.}~\bibnamefont {Johner}}, \
  and\ \bibinfo {author} {\bibfnamefont {A.}~\bibnamefont {Elaissari}},\
  }\bibfield  {title} {\enquote {\bibinfo {title} {Conformation of weakly
  charged polyelectrolytes at a solid--liquid interface},}\ }\href@noop {}
  {\bibfield  {journal} {\bibinfo  {journal} {The Journal of chemical physics}\
  }\textbf {\bibinfo {volume} {94}},\ \bibinfo {pages} {6873--6878} (\bibinfo
  {year} {1991})}\BibitemShut {NoStop}%
\bibitem [{\citenamefont {Varoqui}(1993)}]{varoqui1993structure}%
  \BibitemOpen
  \bibfield  {author} {\bibinfo {author} {\bibfnamefont {R.}~\bibnamefont
  {Varoqui}},\ }\bibfield  {title} {\enquote {\bibinfo {title} {Structure of
  weakly charged polyelectrolytes at a solid-liquid interface},}\ }\href@noop
  {} {\bibfield  {journal} {\bibinfo  {journal} {Journal de Physique II}\
  }\textbf {\bibinfo {volume} {3}},\ \bibinfo {pages} {1097--1108} (\bibinfo
  {year} {1993})}\BibitemShut {NoStop}%
\bibitem [{\citenamefont {Borukhov}, \citenamefont {Andelman},\ and\
  \citenamefont {Orland}(1995)}]{borukhov1995}%
  \BibitemOpen
  \bibfield  {author} {\bibinfo {author} {\bibfnamefont {I.}~\bibnamefont
  {Borukhov}}, \bibinfo {author} {\bibfnamefont {D.}~\bibnamefont {Andelman}},
  \ and\ \bibinfo {author} {\bibfnamefont {H.}~\bibnamefont {Orland}},\
  }\bibfield  {title} {\enquote {\bibinfo {title} {{Polyelectrolyte solutions
  between charged surfaces}},}\ }\href {\doibase 10.1209/0295-5075/32/6/007}
  {\bibfield  {journal} {\bibinfo  {journal} {Epl}\ }\textbf {\bibinfo {volume}
  {32}},\ \bibinfo {pages} {499--504} (\bibinfo {year} {1995})}\BibitemShut
  {NoStop}%
\bibitem [{\citenamefont {Ch{\^a}tellier}\ and\ \citenamefont
  {Joanny}(1996)}]{chatellier1996adsorption}%
  \BibitemOpen
  \bibfield  {author} {\bibinfo {author} {\bibfnamefont {X.}~\bibnamefont
  {Ch{\^a}tellier}}\ and\ \bibinfo {author} {\bibfnamefont {J.-F.}\
  \bibnamefont {Joanny}},\ }\bibfield  {title} {\enquote {\bibinfo {title}
  {Adsorption of polyelectrolyte solutions on surfaces: a debye-h{\"u}ckel
  theory},}\ }\href@noop {} {\bibfield  {journal} {\bibinfo  {journal} {Journal
  de Physique II}\ }\textbf {\bibinfo {volume} {6}},\ \bibinfo {pages}
  {1669--1686} (\bibinfo {year} {1996})}\BibitemShut {NoStop}%
\bibitem [{\citenamefont {Borukhov}, \citenamefont {Andelman},\ and\
  \citenamefont {Orland}(1998)}]{borukhov1998scaling}%
  \BibitemOpen
  \bibfield  {author} {\bibinfo {author} {\bibfnamefont {I.}~\bibnamefont
  {Borukhov}}, \bibinfo {author} {\bibfnamefont {D.}~\bibnamefont {Andelman}},
  \ and\ \bibinfo {author} {\bibfnamefont {H.}~\bibnamefont {Orland}},\
  }\bibfield  {title} {\enquote {\bibinfo {title} {Scaling laws of
  polyelectrolyte adsorption},}\ }\href@noop {} {\bibfield  {journal} {\bibinfo
   {journal} {Macromolecules}\ }\textbf {\bibinfo {volume} {31}},\ \bibinfo
  {pages} {1665--1671} (\bibinfo {year} {1998})}\BibitemShut {NoStop}%
\bibitem [{\citenamefont {Joanny}(1999)}]{joanny1999}%
  \BibitemOpen
  \bibfield  {author} {\bibinfo {author} {\bibfnamefont {J.~F.}\ \bibnamefont
  {Joanny}},\ }\bibfield  {title} {\enquote {\bibinfo {title} {{Polyelectrolyte
  adsorption and charge inversion}},}\ }\href {\doibase 10.1007/s100510050747}
  {\bibfield  {journal} {\bibinfo  {journal} {European Physical Journal B}\
  }\textbf {\bibinfo {volume} {9}},\ \bibinfo {pages} {117--122} (\bibinfo
  {year} {1999})}\BibitemShut {NoStop}%
\bibitem [{\citenamefont {Podgornik}\ and\ \citenamefont
  {J{\"o}nsson}(1993)}]{podgornik1993stretching}%
  \BibitemOpen
  \bibfield  {author} {\bibinfo {author} {\bibfnamefont {R.}~\bibnamefont
  {Podgornik}}\ and\ \bibinfo {author} {\bibfnamefont {B.}~\bibnamefont
  {J{\"o}nsson}},\ }\bibfield  {title} {\enquote {\bibinfo {title} {Stretching
  of polyelectrolyte chains by oppositely charged aggregates},}\ }\href@noop {}
  {\bibfield  {journal} {\bibinfo  {journal} {EPL (Europhysics Letters)}\
  }\textbf {\bibinfo {volume} {24}},\ \bibinfo {pages} {501} (\bibinfo {year}
  {1993})}\BibitemShut {NoStop}%
\bibitem [{\citenamefont {Brilliantov}, \citenamefont {Budkov},\ and\
  \citenamefont {Seidel}(2016)}]{brilliantov2016generation}%
  \BibitemOpen
  \bibfield  {author} {\bibinfo {author} {\bibfnamefont {N.~V.}\ \bibnamefont
  {Brilliantov}}, \bibinfo {author} {\bibfnamefont {Y.~A.}\ \bibnamefont
  {Budkov}}, \ and\ \bibinfo {author} {\bibfnamefont {C.}~\bibnamefont
  {Seidel}},\ }\bibfield  {title} {\enquote {\bibinfo {title} {Generation of
  mechanical force by grafted polyelectrolytes in an electric field},}\
  }\href@noop {} {\bibfield  {journal} {\bibinfo  {journal} {Physical Review
  E}\ }\textbf {\bibinfo {volume} {93}},\ \bibinfo {pages} {032505} (\bibinfo
  {year} {2016})}\BibitemShut {NoStop}%
\bibitem [{\citenamefont {Bohmer}, \citenamefont {Evers},\ and\ \citenamefont
  {Scheutjens}(1990)}]{bohmer1990weak}%
  \BibitemOpen
  \bibfield  {author} {\bibinfo {author} {\bibfnamefont {M.~R.}\ \bibnamefont
  {Bohmer}}, \bibinfo {author} {\bibfnamefont {O.~A.}\ \bibnamefont {Evers}}, \
  and\ \bibinfo {author} {\bibfnamefont {J.~M.}\ \bibnamefont {Scheutjens}},\
  }\bibfield  {title} {\enquote {\bibinfo {title} {Weak polyelectrolytes
  between two surfaces: adsorption and stabilization},}\ }\href@noop {}
  {\bibfield  {journal} {\bibinfo  {journal} {Macromolecules}\ }\textbf
  {\bibinfo {volume} {23}},\ \bibinfo {pages} {2288--2301} (\bibinfo {year}
  {1990})}\BibitemShut {NoStop}%
\bibitem [{\citenamefont {Li}\ and\ \citenamefont
  {Wu}(2006{\natexlab{a}})}]{Li2006}%
  \BibitemOpen
  \bibfield  {author} {\bibinfo {author} {\bibfnamefont {Z.}~\bibnamefont
  {Li}}\ and\ \bibinfo {author} {\bibfnamefont {J.}~\bibnamefont {Wu}},\
  }\bibfield  {title} {\enquote {\bibinfo {title} {{Density functional theory
  for planar electric double layers: Closing the gap between simple and
  polyelectrolytes}},}\ }\href {\doibase 10.1021/jp060127w} {\bibfield
  {journal} {\bibinfo  {journal} {Journal of Physical Chemistry B}\ }\textbf
  {\bibinfo {volume} {110}},\ \bibinfo {pages} {7473--7484} (\bibinfo {year}
  {2006}{\natexlab{a}})}\BibitemShut {NoStop}%
\bibitem [{\citenamefont {Li}\ and\ \citenamefont
  {Wu}(2006{\natexlab{b}})}]{Li2006a}%
  \BibitemOpen
  \bibfield  {author} {\bibinfo {author} {\bibfnamefont {Z.}~\bibnamefont
  {Li}}\ and\ \bibinfo {author} {\bibfnamefont {J.}~\bibnamefont {Wu}},\
  }\bibfield  {title} {\enquote {\bibinfo {title} {{Density functional theory
  for polyelectrolytes near oppositely charged surfaces}},}\ }\href {\doibase
  10.1103/PhysRevLett.96.048302} {\bibfield  {journal} {\bibinfo  {journal}
  {Physical Review Letters}\ }\textbf {\bibinfo {volume} {96}},\ \bibinfo
  {pages} {1--4} (\bibinfo {year} {2006}{\natexlab{b}})}\BibitemShut {NoStop}%
\bibitem [{\citenamefont {Kumar}\ \emph {et~al.}(2017)\citenamefont {Kumar},
  \citenamefont {Mahalik}, \citenamefont {Bocharova}, \citenamefont {Stacy},
  \citenamefont {Gainaru}, \citenamefont {Saito}, \citenamefont {Gobet},
  \citenamefont {Greenbaum}, \citenamefont {Sumpter},\ and\ \citenamefont
  {Sokolov}}]{Kumar2017}%
  \BibitemOpen
  \bibfield  {author} {\bibinfo {author} {\bibfnamefont {R.}~\bibnamefont
  {Kumar}}, \bibinfo {author} {\bibfnamefont {J.~P.}\ \bibnamefont {Mahalik}},
  \bibinfo {author} {\bibfnamefont {V.}~\bibnamefont {Bocharova}}, \bibinfo
  {author} {\bibfnamefont {E.~W.}\ \bibnamefont {Stacy}}, \bibinfo {author}
  {\bibfnamefont {C.}~\bibnamefont {Gainaru}}, \bibinfo {author} {\bibfnamefont
  {T.}~\bibnamefont {Saito}}, \bibinfo {author} {\bibfnamefont {M.~P.}\
  \bibnamefont {Gobet}}, \bibinfo {author} {\bibfnamefont {S.}~\bibnamefont
  {Greenbaum}}, \bibinfo {author} {\bibfnamefont {B.~G.}\ \bibnamefont
  {Sumpter}}, \ and\ \bibinfo {author} {\bibfnamefont {A.~P.}\ \bibnamefont
  {Sokolov}},\ }\bibfield  {title} {\enquote {\bibinfo {title} {{A Rayleighian
  approach for modeling kinetics of ionic transport in polymeric media}},}\
  }\href {\doibase 10.1063/1.4975309} {\bibfield  {journal} {\bibinfo
  {journal} {Journal of Chemical Physics}\ }\textbf {\bibinfo {volume} {146}}
  (\bibinfo {year} {2017}),\ 10.1063/1.4975309}\BibitemShut {NoStop}%
\bibitem [{\citenamefont {Kumar}\ \emph {et~al.}(2020)\citenamefont {Kumar},
  \citenamefont {Mahalik}, \citenamefont {Silmore}, \citenamefont
  {Wojnarowska}, \citenamefont {Erwin}, \citenamefont {Ankner}, \citenamefont
  {Sokolov}, \citenamefont {Sumpter},\ and\ \citenamefont
  {Bocharova}}]{Kumar2020}%
  \BibitemOpen
  \bibfield  {author} {\bibinfo {author} {\bibfnamefont {R.}~\bibnamefont
  {Kumar}}, \bibinfo {author} {\bibfnamefont {J.~P.}\ \bibnamefont {Mahalik}},
  \bibinfo {author} {\bibfnamefont {K.~S.}\ \bibnamefont {Silmore}}, \bibinfo
  {author} {\bibfnamefont {Z.}~\bibnamefont {Wojnarowska}}, \bibinfo {author}
  {\bibfnamefont {A.}~\bibnamefont {Erwin}}, \bibinfo {author} {\bibfnamefont
  {J.~F.}\ \bibnamefont {Ankner}}, \bibinfo {author} {\bibfnamefont {A.~P.}\
  \bibnamefont {Sokolov}}, \bibinfo {author} {\bibfnamefont {B.~G.}\
  \bibnamefont {Sumpter}}, \ and\ \bibinfo {author} {\bibfnamefont
  {V.}~\bibnamefont {Bocharova}},\ }\bibfield  {title} {\enquote {\bibinfo
  {title} {{Capacitance of thin films containing polymerized ionic liquids}},}\
  }\href {\doibase 10.1126/sciadv.aba7952} {\bibfield  {journal} {\bibinfo
  {journal} {Science Advances}\ }\textbf {\bibinfo {volume} {6}} (\bibinfo
  {year} {2020}),\ 10.1126/sciadv.aba7952}\BibitemShut {NoStop}%
\bibitem [{\citenamefont {Kornyshev}(2007)}]{kornyshev2007double}%
  \BibitemOpen
  \bibfield  {author} {\bibinfo {author} {\bibfnamefont {A.~A.}\ \bibnamefont
  {Kornyshev}},\ }\bibfield  {title} {\enquote {\bibinfo {title} {Double-layer
  in ionic liquids: paradigm change?}}\ }\href@noop {} {\bibfield  {journal}
  {\bibinfo  {journal} {The Journal of Physical Chemistry B}\ }\textbf
  {\bibinfo {volume} {111}},\ \bibinfo {pages} {5545--5557} (\bibinfo {year}
  {2007})}\BibitemShut {NoStop}%
\bibitem [{\citenamefont {Budkov}\ \emph {et~al.}(2018)\citenamefont {Budkov},
  \citenamefont {Kolesnikov}, \citenamefont {Goodwin}, \citenamefont
  {Kiselev},\ and\ \citenamefont {Kornyshev}}]{budkov2018theory}%
  \BibitemOpen
  \bibfield  {author} {\bibinfo {author} {\bibfnamefont {Y.~A.}\ \bibnamefont
  {Budkov}}, \bibinfo {author} {\bibfnamefont {A.~L.}\ \bibnamefont
  {Kolesnikov}}, \bibinfo {author} {\bibfnamefont {Z.~A.}\ \bibnamefont
  {Goodwin}}, \bibinfo {author} {\bibfnamefont {M.~G.}\ \bibnamefont
  {Kiselev}}, \ and\ \bibinfo {author} {\bibfnamefont {A.~A.}\ \bibnamefont
  {Kornyshev}},\ }\bibfield  {title} {\enquote {\bibinfo {title} {Theory of
  electrosorption of water from ionic liquids},}\ }\href@noop {} {\bibfield
  {journal} {\bibinfo  {journal} {Electrochimica Acta}\ }\textbf {\bibinfo
  {volume} {284}},\ \bibinfo {pages} {346--354} (\bibinfo {year}
  {2018})}\BibitemShut {NoStop}%
\bibitem [{\citenamefont {Wu}\ \emph {et~al.}(2011)\citenamefont {Wu},
  \citenamefont {Jiang}, \citenamefont {Jiang}, \citenamefont {Jin},\ and\
  \citenamefont {Henderson}}]{wu2011classical}%
  \BibitemOpen
  \bibfield  {author} {\bibinfo {author} {\bibfnamefont {J.}~\bibnamefont
  {Wu}}, \bibinfo {author} {\bibfnamefont {T.}~\bibnamefont {Jiang}}, \bibinfo
  {author} {\bibfnamefont {D.}~\bibnamefont {Jiang}}, \bibinfo {author}
  {\bibfnamefont {Z.}~\bibnamefont {Jin}}, \ and\ \bibinfo {author}
  {\bibfnamefont {D.}~\bibnamefont {Henderson}},\ }\bibfield  {title} {\enquote
  {\bibinfo {title} {A classical density functional theory for interfacial
  layering of ionic liquids},}\ }\href@noop {} {\bibfield  {journal} {\bibinfo
  {journal} {Soft Matter}\ }\textbf {\bibinfo {volume} {7}},\ \bibinfo {pages}
  {11222--11231} (\bibinfo {year} {2011})}\BibitemShut {NoStop}%
\bibitem [{\citenamefont {Jiang}, \citenamefont {Meng},\ and\ \citenamefont
  {Wu}(2011)}]{jiang2011density}%
  \BibitemOpen
  \bibfield  {author} {\bibinfo {author} {\bibfnamefont {D.}~\bibnamefont
  {Jiang}}, \bibinfo {author} {\bibfnamefont {D.}~\bibnamefont {Meng}}, \ and\
  \bibinfo {author} {\bibfnamefont {J.}~\bibnamefont {Wu}},\ }\bibfield
  {title} {\enquote {\bibinfo {title} {Density functional theory for
  differential capacitance of planar electric double layers in ionic
  liquids},}\ }\href@noop {} {\bibfield  {journal} {\bibinfo  {journal}
  {Chemical Physics Letters}\ }\textbf {\bibinfo {volume} {504}},\ \bibinfo
  {pages} {153--158} (\bibinfo {year} {2011})}\BibitemShut {NoStop}%
\bibitem [{\citenamefont {Forsman}, \citenamefont {Woodward},\ and\
  \citenamefont {Trulsson}(2011)}]{forsman2011classical}%
  \BibitemOpen
  \bibfield  {author} {\bibinfo {author} {\bibfnamefont {J.}~\bibnamefont
  {Forsman}}, \bibinfo {author} {\bibfnamefont {C.~E.}\ \bibnamefont
  {Woodward}}, \ and\ \bibinfo {author} {\bibfnamefont {M.}~\bibnamefont
  {Trulsson}},\ }\bibfield  {title} {\enquote {\bibinfo {title} {A classical
  density functional theory of ionic liquids},}\ }\href@noop {} {\bibfield
  {journal} {\bibinfo  {journal} {The Journal of Physical Chemistry B}\
  }\textbf {\bibinfo {volume} {115}},\ \bibinfo {pages} {4606--4612} (\bibinfo
  {year} {2011})}\BibitemShut {NoStop}%
\bibitem [{\citenamefont {Goodwin}, \citenamefont {Feng},\ and\ \citenamefont
  {Kornyshev}(2017)}]{goodwin2017mean}%
  \BibitemOpen
  \bibfield  {author} {\bibinfo {author} {\bibfnamefont {Z.~A.}\ \bibnamefont
  {Goodwin}}, \bibinfo {author} {\bibfnamefont {G.}~\bibnamefont {Feng}}, \
  and\ \bibinfo {author} {\bibfnamefont {A.~A.}\ \bibnamefont {Kornyshev}},\
  }\bibfield  {title} {\enquote {\bibinfo {title} {Mean-field theory of
  electrical double layer in ionic liquids with account of short-range
  correlations},}\ }\href@noop {} {\bibfield  {journal} {\bibinfo  {journal}
  {Electrochimica Acta}\ }\textbf {\bibinfo {volume} {225}},\ \bibinfo {pages}
  {190--197} (\bibinfo {year} {2017})}\BibitemShut {NoStop}%
\bibitem [{\citenamefont {Maggs}\ and\ \citenamefont
  {Podgornik}(2016)}]{Maggs2016}%
  \BibitemOpen
  \bibfield  {author} {\bibinfo {author} {\bibfnamefont {A.~C.}\ \bibnamefont
  {Maggs}}\ and\ \bibinfo {author} {\bibfnamefont {R.}~\bibnamefont
  {Podgornik}},\ }\bibfield  {title} {\enquote {\bibinfo {title} {{General
  theory of asymmetric steric interactions in electrostatic double layers}},}\
  }\href {\doibase 10.1039/c5sm01757b} {\bibfield  {journal} {\bibinfo
  {journal} {Soft Matter}\ }\textbf {\bibinfo {volume} {12}},\ \bibinfo {pages}
  {1219--1229} (\bibinfo {year} {2016})}\BibitemShut {NoStop}%
\bibitem [{\citenamefont {May}(2019)}]{may2019differential}%
  \BibitemOpen
  \bibfield  {author} {\bibinfo {author} {\bibfnamefont {S.}~\bibnamefont
  {May}},\ }\bibfield  {title} {\enquote {\bibinfo {title} {Differential
  capacitance of the electric double layer: mean-field modeling approaches},}\
  }\href@noop {} {\bibfield  {journal} {\bibinfo  {journal} {Current Opinion in
  Electrochemistry}\ }\textbf {\bibinfo {volume} {13}},\ \bibinfo {pages}
  {125--131} (\bibinfo {year} {2019})}\BibitemShut {NoStop}%
\bibitem [{\citenamefont {Cruz}\ \emph {et~al.}(2019)\citenamefont {Cruz},
  \citenamefont {Kondrat}, \citenamefont {Lomba},\ and\ \citenamefont
  {Ciach}}]{cruz2019effect}%
  \BibitemOpen
  \bibfield  {author} {\bibinfo {author} {\bibfnamefont {C.}~\bibnamefont
  {Cruz}}, \bibinfo {author} {\bibfnamefont {S.}~\bibnamefont {Kondrat}},
  \bibinfo {author} {\bibfnamefont {E.}~\bibnamefont {Lomba}}, \ and\ \bibinfo
  {author} {\bibfnamefont {A.}~\bibnamefont {Ciach}},\ }\bibfield  {title}
  {\enquote {\bibinfo {title} {Effect of proximity to ionic liquid-solvent
  demixing on electrical double layers},}\ }\href@noop {} {\bibfield  {journal}
  {\bibinfo  {journal} {Journal of Molecular Liquids}\ }\textbf {\bibinfo
  {volume} {294}},\ \bibinfo {pages} {111368} (\bibinfo {year}
  {2019})}\BibitemShut {NoStop}%
\bibitem [{\citenamefont {Bazant}, \citenamefont {Storey},\ and\ \citenamefont
  {Kornyshev}(2011)}]{bazant2011double}%
  \BibitemOpen
  \bibfield  {author} {\bibinfo {author} {\bibfnamefont {M.~Z.}\ \bibnamefont
  {Bazant}}, \bibinfo {author} {\bibfnamefont {B.~D.}\ \bibnamefont {Storey}},
  \ and\ \bibinfo {author} {\bibfnamefont {A.~A.}\ \bibnamefont {Kornyshev}},\
  }\bibfield  {title} {\enquote {\bibinfo {title} {Double layer in ionic
  liquids: Overscreening versus crowding},}\ }\href@noop {} {\bibfield
  {journal} {\bibinfo  {journal} {Physical review letters}\ }\textbf {\bibinfo
  {volume} {106}},\ \bibinfo {pages} {046102} (\bibinfo {year}
  {2011})}\BibitemShut {NoStop}%
\bibitem [{\citenamefont {Avni}, \citenamefont {Adar},\ and\ \citenamefont
  {Andelman}(2020)}]{avni2020charge}%
  \BibitemOpen
  \bibfield  {author} {\bibinfo {author} {\bibfnamefont {Y.}~\bibnamefont
  {Avni}}, \bibinfo {author} {\bibfnamefont {R.~M.}\ \bibnamefont {Adar}}, \
  and\ \bibinfo {author} {\bibfnamefont {D.}~\bibnamefont {Andelman}},\
  }\bibfield  {title} {\enquote {\bibinfo {title} {Charge oscillations in ionic
  liquids: A microscopic cluster model},}\ }\href@noop {} {\bibfield  {journal}
  {\bibinfo  {journal} {Physical Review E}\ }\textbf {\bibinfo {volume}
  {101}},\ \bibinfo {pages} {010601} (\bibinfo {year} {2020})}\BibitemShut
  {NoStop}%
\bibitem [{\citenamefont {Budkov}, \citenamefont {Zavarzin},\ and\
  \citenamefont {Kolesnikov}(2021)}]{budkovJPCC2021}%
  \BibitemOpen
  \bibfield  {author} {\bibinfo {author} {\bibfnamefont {Y.~A.}\ \bibnamefont
  {Budkov}}, \bibinfo {author} {\bibfnamefont {S.~V.}\ \bibnamefont
  {Zavarzin}}, \ and\ \bibinfo {author} {\bibfnamefont {A.~L.}\ \bibnamefont
  {Kolesnikov}},\ }\bibfield  {title} {\enquote {\bibinfo {title} {Theory of
  ionic liquids with polarizable ions on a charged electrode},}\ }\href@noop {}
  {\ \textbf {\bibinfo {volume} {125}},\ \bibinfo {pages} {21151--21159}
  (\bibinfo {year} {2021})}\BibitemShut {NoStop}%
\bibitem [{\citenamefont {Lifshitz}(1969)}]{lifshitz1969some}%
  \BibitemOpen
  \bibfield  {author} {\bibinfo {author} {\bibfnamefont {I.}~\bibnamefont
  {Lifshitz}},\ }\bibfield  {title} {\enquote {\bibinfo {title} {Some problems
  of the statistical theory of biopolymers},}\ }\href@noop {} {\bibfield
  {journal} {\bibinfo  {journal} {Sov. Phys. JETP}\ }\textbf {\bibinfo {volume}
  {28}},\ \bibinfo {pages} {1280--1286} (\bibinfo {year} {1969})}\BibitemShut
  {NoStop}%
\bibitem [{\citenamefont {Grosberg}\ and\ \citenamefont
  {Khokhlov}(1994)}]{khokhlov1994statistical}%
  \BibitemOpen
  \bibfield  {author} {\bibinfo {author} {\bibfnamefont {A.~Y.}\ \bibnamefont
  {Grosberg}}\ and\ \bibinfo {author} {\bibfnamefont {A.~R.}\ \bibnamefont
  {Khokhlov}},\ }\href@noop {} {\emph {\bibinfo {title} {Statistical physics of
  macromolecules}}}\ (\bibinfo  {publisher} {Amer Inst of Physics},\ \bibinfo
  {year} {1994})\BibitemShut {NoStop}%
\bibitem [{\citenamefont {Sanchez}\ and\ \citenamefont
  {Lacombe}(1978)}]{sanchez1978statistical}%
  \BibitemOpen
  \bibfield  {author} {\bibinfo {author} {\bibfnamefont {I.~C.}\ \bibnamefont
  {Sanchez}}\ and\ \bibinfo {author} {\bibfnamefont {R.~H.}\ \bibnamefont
  {Lacombe}},\ }\bibfield  {title} {\enquote {\bibinfo {title} {Statistical
  thermodynamics of polymer solutions},}\ }\href@noop {} {\bibfield  {journal}
  {\bibinfo  {journal} {Macromolecules}\ }\textbf {\bibinfo {volume} {11}},\
  \bibinfo {pages} {1145--1156} (\bibinfo {year} {1978})}\BibitemShut {NoStop}%
\bibitem [{\citenamefont {Borue}\ and\ \citenamefont
  {Erukhimovich}(1988)}]{Borue1988}%
  \BibitemOpen
  \bibfield  {author} {\bibinfo {author} {\bibfnamefont {V.~Y.}\ \bibnamefont
  {Borue}}\ and\ \bibinfo {author} {\bibfnamefont {I.~Y.}\ \bibnamefont
  {Erukhimovich}},\ }\bibfield  {title} {\enquote {\bibinfo {title} {{A
  Statistical Theory of Weakly Charged Polyelectrolytes: Fluctuations, Equation
  of State and Microphase Separation}},}\ }\href {\doibase 10.1021/ma00189a019}
  {\bibfield  {journal} {\bibinfo  {journal} {Macromolecules}\ }\textbf
  {\bibinfo {volume} {21}},\ \bibinfo {pages} {3240--3249} (\bibinfo {year}
  {1988})}\BibitemShut {NoStop}%
\bibitem [{\citenamefont {Muthukumar}(1996)}]{muthukumar1996double}%
  \BibitemOpen
  \bibfield  {author} {\bibinfo {author} {\bibfnamefont {M.}~\bibnamefont
  {Muthukumar}},\ }\bibfield  {title} {\enquote {\bibinfo {title} {Double
  screening in polyelectrolyte solutions: Limiting laws and crossover
  formulas},}\ }\href@noop {} {\bibfield  {journal} {\bibinfo  {journal} {The
  Journal of chemical physics}\ }\textbf {\bibinfo {volume} {105}},\ \bibinfo
  {pages} {5183--5199} (\bibinfo {year} {1996})}\BibitemShut {NoStop}%
\bibitem [{\citenamefont {Budkov}(2020)}]{budkov2020statistical}%
  \BibitemOpen
  \bibfield  {author} {\bibinfo {author} {\bibfnamefont {Y.~A.}\ \bibnamefont
  {Budkov}},\ }\bibfield  {title} {\enquote {\bibinfo {title} {Statistical
  field theory of ion--molecular solutions},}\ }\href@noop {} {\bibfield
  {journal} {\bibinfo  {journal} {Physical Chemistry Chemical Physics}\
  }\textbf {\bibinfo {volume} {22}},\ \bibinfo {pages} {14756--14772} (\bibinfo
  {year} {2020})}\BibitemShut {NoStop}%
\bibitem [{\citenamefont {Budkov}(2019)}]{budkov2019statistical}%
  \BibitemOpen
  \bibfield  {author} {\bibinfo {author} {\bibfnamefont {Y.~A.}\ \bibnamefont
  {Budkov}},\ }\bibfield  {title} {\enquote {\bibinfo {title} {A statistical
  field theory of salt solutions of "hairy" dielectric particles},}\
  }\href@noop {} {\bibfield  {journal} {\bibinfo  {journal} {Journal of
  Physics: Condensed Matter}\ }\textbf {\bibinfo {volume} {32}},\ \bibinfo
  {pages} {055101} (\bibinfo {year} {2019})}\BibitemShut {NoStop}%
\bibitem [{\citenamefont {Slavchov}(2014)}]{slavchov2014quadrupole}%
  \BibitemOpen
  \bibfield  {author} {\bibinfo {author} {\bibfnamefont {R.~I.}\ \bibnamefont
  {Slavchov}},\ }\bibfield  {title} {\enquote {\bibinfo {title} {Quadrupole
  terms in the maxwell equations: Debye-h{\"u}ckel theory in quadrupolarizable
  solvent and self-salting-out of electrolytes},}\ }\href@noop {} {\bibfield
  {journal} {\bibinfo  {journal} {The Journal of Chemical Physics}\ }\textbf
  {\bibinfo {volume} {140}},\ \bibinfo {pages} {074503} (\bibinfo {year}
  {2014})}\BibitemShut {NoStop}%
\bibitem [{\citenamefont {Andelman}\ and\ \citenamefont
  {Joanny}(2000{\natexlab{b}})}]{Andelman2000}%
  \BibitemOpen
  \bibfield  {author} {\bibinfo {author} {\bibfnamefont {D.}~\bibnamefont
  {Andelman}}\ and\ \bibinfo {author} {\bibfnamefont {J.~F.}\ \bibnamefont
  {Joanny}},\ }\bibfield  {title} {\enquote {\bibinfo {title} {{Polyelectrolyte
  adsorption}},}\ }\href {\doibase 10.1016/S1296-2147(00)01130-6} {\bibfield
  {journal} {\bibinfo  {journal} {Comptes Rendus de l'Academie des Sciences -
  Series IV: Physics, Astrophysics}\ }\textbf {\bibinfo {volume} {1}},\
  \bibinfo {pages} {1153--1162} (\bibinfo {year} {2000}{\natexlab{b}})},\
  \Eprint {http://arxiv.org/abs/0011072} {arXiv:0011072 [cond-mat]}
  \BibitemShut {NoStop}%
\bibitem [{\citenamefont {Boubl{\'\i}k}(1970)}]{boublik1970hard}%
  \BibitemOpen
  \bibfield  {author} {\bibinfo {author} {\bibfnamefont {T.}~\bibnamefont
  {Boubl{\'\i}k}},\ }\bibfield  {title} {\enquote {\bibinfo {title}
  {Hard-sphere equation of state},}\ }\href@noop {} {\bibfield  {journal}
  {\bibinfo  {journal} {The Journal of chemical physics}\ }\textbf {\bibinfo
  {volume} {53}},\ \bibinfo {pages} {471--472} (\bibinfo {year}
  {1970})}\BibitemShut {NoStop}%
\bibitem [{\citenamefont {Mansoori}\ \emph {et~al.}(1971)\citenamefont
  {Mansoori}, \citenamefont {Carnahan}, \citenamefont {Starling},\ and\
  \citenamefont {Leland~Jr}}]{mansoori1971equilibrium}%
  \BibitemOpen
  \bibfield  {author} {\bibinfo {author} {\bibfnamefont {G.}~\bibnamefont
  {Mansoori}}, \bibinfo {author} {\bibfnamefont {N.~F.}\ \bibnamefont
  {Carnahan}}, \bibinfo {author} {\bibfnamefont {K.}~\bibnamefont {Starling}},
  \ and\ \bibinfo {author} {\bibfnamefont {T.}~\bibnamefont {Leland~Jr}},\
  }\bibfield  {title} {\enquote {\bibinfo {title} {Equilibrium thermodynamic
  properties of the mixture of hard spheres},}\ }\href@noop {} {\bibfield
  {journal} {\bibinfo  {journal} {The Journal of Chemical Physics}\ }\textbf
  {\bibinfo {volume} {54}},\ \bibinfo {pages} {1523--1525} (\bibinfo {year}
  {1971})}\BibitemShut {NoStop}%
\bibitem [{\citenamefont {Budkov}\ \emph {et~al.}(2020)\citenamefont {Budkov},
  \citenamefont {Sergeev}, \citenamefont {Zavarzin},\ and\ \citenamefont
  {Kolesnikov}}]{budkov2020two}%
  \BibitemOpen
  \bibfield  {author} {\bibinfo {author} {\bibfnamefont {Y.~A.}\ \bibnamefont
  {Budkov}}, \bibinfo {author} {\bibfnamefont {A.~V.}\ \bibnamefont {Sergeev}},
  \bibinfo {author} {\bibfnamefont {S.~V.}\ \bibnamefont {Zavarzin}}, \ and\
  \bibinfo {author} {\bibfnamefont {A.~L.}\ \bibnamefont {Kolesnikov}},\
  }\bibfield  {title} {\enquote {\bibinfo {title} {Two-component electrolyte
  solutions with dipolar cations on a charged electrode: Theory and computer
  simulations},}\ }\href {\doibase 10.1021/acs.jpcc.0c03623} {\bibfield
  {journal} {\bibinfo  {journal} {The Journal of Physical Chemistry C}\
  }\textbf {\bibinfo {volume} {124}},\ \bibinfo {pages} {16308--16314}
  (\bibinfo {year} {2020})}\BibitemShut {NoStop}%
\bibitem [{\citenamefont {Budkov}, \citenamefont {Kolesnikov},\ and\
  \citenamefont {Kiselev}(2015)}]{budkov2015modified}%
  \BibitemOpen
  \bibfield  {author} {\bibinfo {author} {\bibfnamefont {Y.~A.}\ \bibnamefont
  {Budkov}}, \bibinfo {author} {\bibfnamefont {A.}~\bibnamefont {Kolesnikov}},
  \ and\ \bibinfo {author} {\bibfnamefont {M.}~\bibnamefont {Kiselev}},\
  }\bibfield  {title} {\enquote {\bibinfo {title} {A modified poisson-boltzmann
  theory: Effects of co-solvent polarizability},}\ }\href@noop {} {\bibfield
  {journal} {\bibinfo  {journal} {EPL (Europhysics Letters)}\ }\textbf
  {\bibinfo {volume} {111}},\ \bibinfo {pages} {28002} (\bibinfo {year}
  {2015})}\BibitemShut {NoStop}%
\bibitem [{\citenamefont {Budkov}, \citenamefont {Kolesnikov},\ and\
  \citenamefont {Kiselev}(2016)}]{budkov2016theory}%
  \BibitemOpen
  \bibfield  {author} {\bibinfo {author} {\bibfnamefont {Y.~A.}\ \bibnamefont
  {Budkov}}, \bibinfo {author} {\bibfnamefont {A.}~\bibnamefont {Kolesnikov}},
  \ and\ \bibinfo {author} {\bibfnamefont {M.}~\bibnamefont {Kiselev}},\
  }\bibfield  {title} {\enquote {\bibinfo {title} {On the theory of electric
  double layer with explicit account of a polarizable co-solvent},}\
  }\href@noop {} {\bibfield  {journal} {\bibinfo  {journal} {The Journal of
  Chemical Physics}\ }\textbf {\bibinfo {volume} {144}},\ \bibinfo {pages}
  {184703} (\bibinfo {year} {2016})}\BibitemShut {NoStop}%
\bibitem [{\citenamefont {Abrashkin}, \citenamefont {Andelman},\ and\
  \citenamefont {Orland}(2007)}]{abrashkin2007dipolar}%
  \BibitemOpen
  \bibfield  {author} {\bibinfo {author} {\bibfnamefont {A.}~\bibnamefont
  {Abrashkin}}, \bibinfo {author} {\bibfnamefont {D.}~\bibnamefont {Andelman}},
  \ and\ \bibinfo {author} {\bibfnamefont {H.}~\bibnamefont {Orland}},\
  }\bibfield  {title} {\enquote {\bibinfo {title} {Dipolar poisson-boltzmann
  equation: ions and dipoles close to charge interfaces},}\ }\href@noop {}
  {\bibfield  {journal} {\bibinfo  {journal} {Physical review letters}\
  }\textbf {\bibinfo {volume} {99}},\ \bibinfo {pages} {077801} (\bibinfo
  {year} {2007})}\BibitemShut {NoStop}%
\end{thebibliography}%
\end{document}